\documentclass[12pt]{article}

\usepackage{geometry}
\usepackage{amsfonts,amsmath,amssymb,amsthm}
\usepackage{mathrsfs}
\usepackage{lmodern}
\usepackage{graphicx,subcaption}
\usepackage[colorlinks,citecolor=blue,linkcolor=red,bookmarks=false,hypertexnames=true]{hyperref}
\usepackage[font=footnotesize,labelfont={bf},labelsep=period]{caption}
\usepackage[table]{xcolor}
\usepackage{lettrine}

\usepackage{natbib}
\begin{document}

\title{Transition path theory insights into hurricane rapid intensification}

\author{
F.J.\ Beron-Vera\thanks{Corresponding author.}\\ Department of Atmospheric Sciences\\ Rosenstiel School of Marine, Atmospheric \& Earth Science\\ University of Miami\\ Miami, Florida, USA\\ \href{mailto:fberon@miami.edu}{\texttt{fberon@miami.edu}}
\and
G.\ Bonner\\ Morgridge Institute for Research\\ University of Wisconsin\\ Madison, Wisconsin, USA\\ \href{mailto:gbonner@morgridge.org}{\texttt{gbonner@morgridge.org}}
\and
M.J.\ Olascoaga\\ Department of Ocean Sciences\\ Rosenstiel School of Marine, Atmospheric \& Earth Science\\ University of Miami\\ Miami, Florida, USA\\ \href{mailto:jolascoaga@miami.edu}{\texttt{jolascoaga@miami.edu}}
\and
S.\ Dong\\ Physical Oceanography Division\\ Atlantic Oceanic and Meteorological Laboratory\\ National Oceanic and Atmospheric Administration\\ Miami, Florida, USA\\ \href{mailto:shen.fu@noaal.gov}{\texttt{Shenfu.Dong@noaa.gov}}
\and
H.\ Lopez\\ Physical Oceanography Division\\ Atlantic Oceanic and Meteorological Laboratory\\ National Oceanic and Atmospheric Administration\\ Miami, Florida, USA\\ \href{mailto:hosemay.lopez@noaal.gov}{\texttt{Hosmay.Lopez@noaa.gov}}}

\date{Started: July 1, 2024; this version: \today.}%

\maketitle

\begin{abstract}
    We explore hurricane and ocean reanalysis data to understand how rapid intensification (RI) of tropical cyclones is impacted by the upper ocean density structure, with an emphasis on barrier layer (BL) thickness and thermocline depth in the eastern Caribbean Sea and adjacent western tropical North Atlantic. This analysis leverages transition path theory (TPT), supported by basic statistical methods. In TPT, Markov chains are constructed by discretizing data series related to weather system intensity, changes in intensity, translational speed, and BL thickness and thermocline depth. These series are viewed as trajectories in abstract state spaces, following a memoryless stochastic process. RI imminence is rigorously framed using a newly derived TPT statistic, which gives the time distribution to first reach a target---the RI state---from a source---for instance, the state determined by a certain BL range and system intensity---conditional on connecting paths exhibiting minimal detours. Increased RI frequency is observed in the eastern Caribbean and nearby Atlantic, influenced by river runoff, primarily in tropical storms and category 2 hurricanes. RI frequently correlates with a well-developed BL; however, increased translational speed is necessary for RI. TPT shows a stronger connection between RI and thermocline depth than BL presence, with RI likelihood rising for hurricanes with a thin BL, especially category 1. Across all strength categories, a deep thermocline consistently elevates RI probability, a factor missed by basic statistical analysis. Furthermore, translational speed is crucial, with faster, stronger hurricanes more susceptible to RI, while slower systems are less so.

\end{abstract}


\section{Introduction}

The term \emph{barrier layer} (BL) refers to a layer of comparatively low salinity water found beneath a fresher surface layer, primarily induced by freshwater influx from precipitation or river discharge \citep{Godfrey-Lindstrom-89}. A BL induces pronounced and stable stratification, which effectively impedes vertical mixing processes. In the eastern Caribbean Sea and the adjacent western tropical north Atlantic region, the occurrence of semipermanent BLs is largely attributed to freshwater contributions from the Orinoco and Amazon Rivers \citep{Muller-etal-88, Lentz-95, Corredor-Morell-01, Mignot-etal-07}. This region is on the path of tropical cyclones, which often undergo \emph{rapid intensification} (RI), defined as an increase in intensity (maximum sustained winds) of 30 kt (1 kt = 1.8 km h$^{-1}$) within a day \citep{Garner-etal-23}.  The thermal structure of the ocean plays an essential role in the evolution of these storms \citep{Shay-etal-00, Goni-Trinanes-03, Halliwell-etal-15}, and various studies \citep{Seroka-etal-17, Hlywiak-Nolan-19, Domingues-etal-21} suggest that the salinity structure, through the properties of BL, is expected to impact their RI.

Strong stratification associated with the BL may limit the cooling effect of upwelled cold water \citep{Androulidakis-etal-16, Grodsky-etal-12}, potentially allowing hurricanes to maintain or even increase their intensity over time. This occurs because the BL tends to trap warmer water, which can provide additional heat content to fuel hurricanes as they pass through. Following this reasoning, the thickness of the BL should play an important role in RI. Additional factors like the depth of the thermocline must not be ignored, nor the intensity and speed of the weather systems.

The hypothesis that the BL may influence the RI of tropical cyclones is not new and has been explored in various studies, particularly in the tropical North Pacific \citep{Wang-etal-11} and the Indian Ocean \citep{Neetu-etal-12}. Some research \citep{Balaguru-etal-12} suggests that the BL is favorable to RI, while other studies \citep{Hernandez-etal-16} argue that it has minimal influence based on the analysis of a regional ocean numerical simulation forced with realistic tropical cyclone winds. In the western tropical North Atlantic, research focusing on individual cases \citep{Domingues-etal-15, Rudzin-etal-19} also provides valuable insight. Closer to our goal here, \citet{Balaguru-etal-20}, using an ocean reanalysis product, demonstrate that both the upper ocean thermal structure and sea surface salinity play crucial roles in the RI process.

This paper explores the impact of the density structure of the upper ocean on RI using a combination of standard statistical methods and \emph{transition path theory} (TPT) \citep{VandenEijnden-06, E-VandenEijnden-06}, all applied on historical hurricane and ocean reanalysis data in the eastern Caribbean Sea and the nearby western tropical North Atlantic. A practical framework for applying TPT is provided by a Markov chain \citep{Metzner-etal-09}, which is a stochastic model in which the next state of a dynamical system depends only on the current state, not the entire past. In our case, an appropriate Markov chain describes evolution in an abstract state space with coordinates chosen to be a cross section of relevant data, namely, weather system intensity, intensity change, translational speed, and along-system-path BL thickness and thermocline depth. Various chains are constructed, each by discretizing trajectories in the corresponding state space. A novel TPT statistic is formulated to rigorously characterize RI imminence. With a target representing the RI state, and a source defined by a given initial condition, for instance, characterized by certain BL range or thermocline depth and system intensity, this newly developed TPT statistic provides the distribution of time required to first reach the target from the source, conditional on the connecting paths showing minimal detours.

Our findings indicate that while RI often links with a well-developed BL, faster translational speed is needed for intensifying systems. TPT analysis highlights a stronger connection between RI and thermocline depth than BL presence, with a thin BL increasing RI likelihood. A deep thermocline consistently boosts RI probability across all strength categories, an aspect overlooked by basic statistical methods. Faster, stronger hurricanes are more susceptible to RI, whereas slower ones are less so. Although RI typically ties to a thick BL and deep thermocline, frequent cases with a thin BL challenge initial assumptions.

The remainder of the paper is organized as follows. In Section \ref{sec:data}, we present the main datasets employed in the study (Section \ref{sec:main}) and define the time series on which the analysis will be carried out, recalling definitions and establishing parameter settings (Section \ref{sec:prep}). The results of the analysis are presented in Section \ref{sec:resuls}, with Section \ref{sec:standard} devoted to a straightforward analysis using conventional statistical tools, and Section \ref{sec:tpt} to the specialized TPT analysis. A summary and conclusions are offered in Section \ref{sec:conclusions}. Appendix \ref{sec:tpt-app} reviews details of TPT, its main statistics including the remaining transition time, and the extension of TPT to open dynamical systems. Appendix \ref{sec:tpt-prob} discusses how TPT is applied to the RI problem of interest here. Finally, Appendix \ref{sec:cdf-computation} presents a derivation of a new TPT statistic, the distribution of transition times, which we use to characterize RI imminence.

\section{Data}\label{sec:data}

\subsection{Main datasets}\label{sec:main}

The weather system dataset used in this study is generated by the National Hurricane Center (NHC) and is distributed by the National Oceanic and Atmospheric Administration (NOAA) \citep{Landsea-Franklin-13}.  The specific dataset examined is the Atlantic HURDAT2, which comprises 6-hour data on the location, maximum winds, central pressure, and size of all documented tropical and subtropical weather systems that eventually acquired hurricane classification. The data span the years 1851 to 2023, except for the size, which was recorded only starting in 2004. In this analysis, we have elected to consider data beginning in 1980, marking the onset of the satellite era.

The temperature and salinity fields used in our analyses are derived from the Ocean Reanalysis System 5 (ORAS5) developed by the European Centre for Medium-Range Weather Forecasts (ECMWF). ECMWF evaluates the state of the global ocean through its operational system, OCEAN5 \citep{Zuo-etal-19}. The OCEAN5 framework is a global eddy-permitting ocean-sea ice ensemble reanalysis--analysis system comprised of five members. This system incorporates a behind-real-time component that facilitates the production of ORAS5. Building on its predecessor ORA5P \citep{Zuo-etal-15}, ORAS5 offers a daily estimate of the historical state of the ocean from 1979 to the present, spanning 75 vertical levels. For the purposes of this study, we have chosen to consider temperature and salinity fields within the spatial domain defined by [100$^\circ$W,40$^\circ$W] $\times$ [0$^\circ$,40$^\circ$N], corresponding to the temporal range 1980--2023, which aligns with the period of the hurricane data.

\subsection{Preliminary data preparation}\label{sec:prep}

To prepare for the various analyses, we began by computing \emph{BL thickness} (BLT) as \citep{deBoyer-etal-07}
\begin{equation}
    \text{BLT} = \text{ILD} - \text{MLD}. 
\end{equation}
Here, ILD (\emph{isothermal layer depth}) is the depth where the temperature has decreased by 0.5$^\circ$C from a 5-m reference depth and MLD (mixed-layer depth) is the depth where the density has increased by 0.125 kg\,m$^{-3}$ relative to surface density. Specifics pertaining to the selection of thresholds are detailed in \citet{Zhang-etal-22}. The density profiles are calculated from the temperature and salinity profiles according to the International Thermodynamic Equation of Seawater - 2010 (TEOS10) \citep{Millero-etal-08} as implemented in the Gibbs-SeaWater (GSW) Oceanographic Toolbox \citep{McDougal-Barker-11}. 

The various hydrographic variables were subsequently linearly interpolated along the trajectories of the weather systems. Furthermore, we computed time series of the variation of Intensity---maximum wind sustained by a system along its track---over a 24-hour period, which we loosely term Acceleration. This resulted in five primary and distinct time series for examination: BLT($t$), IDL($t$), Intensity($t$), Acceleration($t$), and Speed($t$), with the latter referring to the translational speed of the system. Each time series consists of data points sampled regularly at 6-hour intervals.

Sea surface salinity (SSS) is excluded from the analysis because it closely correlates with BLT($t$) concerning RI, as will be demonstrated. Thus, it is unnecessary to consider this separately.

RI is defined by the condition $\text{Acceleration} > 30$\,kt/day, which corresponds to an increase in the intensity of the weather system by 30\,kt within one day, as per the convention \citep{Garner-etal-23}. Consequently, the Acceleration($t$) time series is transformed into a binary sequence comprising zeros during periods lacking RI and ones during phases of active RI.

We have chosen to propose the existence or significant formation of a BL when $\text{BLT} > 5$ m. This choice ensures that small measurement errors are unlikely to accumulate and suggest the presence of a BL when none exists. Variations around this threshold do not impact the results.

In the following sections, weather systems are classified according to the intensity achieved along their trajectories, as indicated in Table \ref{tab:class} according to the conventionally applied Saffir--Simpson scale \citep{Kantha-06}.  

\begin{table}[t!]
    \centering
    \begin{tabular}{lll}
         \hline\hline
         Weather system &  Acronyms & Intensity [kt]\\
         Tropical depression & TD, D & $\leq33$\\
         Tropical storm & TS, S & 34--63\\
         Category 1 hurricane & Cat 1, 1 & 64--82\\
         Category 2 hurricane & Cat 2, 2 & 83--95\\
         Category 3 hurricane & Cat 3, 3 & 96--112\\
         Category 4 hurricane & Cat 4, 4 & 113--136\\
         Category 5 hurricane & Cat 5, 5 & $\geq137$\\
         \hline
    \end{tabular}
    \caption{Classification of weather systems according to intensity.}
    \label{tab:class}
\end{table}

\section{Results}\label{sec:resuls}

\subsection{Straightforward statistical analysis}\label{sec:standard}

\begin{figure}
    \centering
    \begin{subfigure}{.49\textwidth}
        \includegraphics[width=\textwidth]{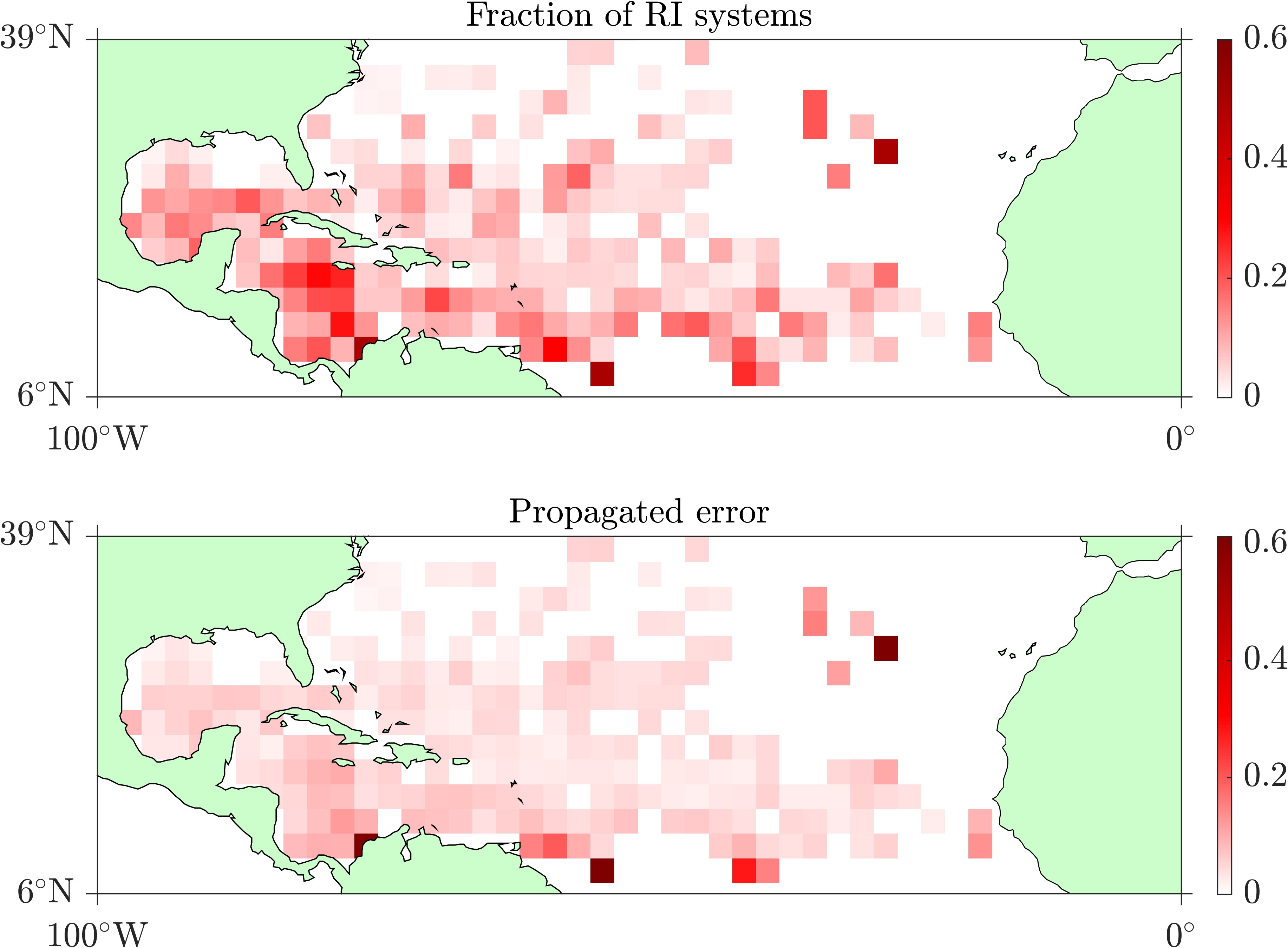}
        \caption{}
    \end{subfigure}
    \begin{subfigure}{.49\textwidth}
        \raisebox{.5cm}{\includegraphics[width=\textwidth]{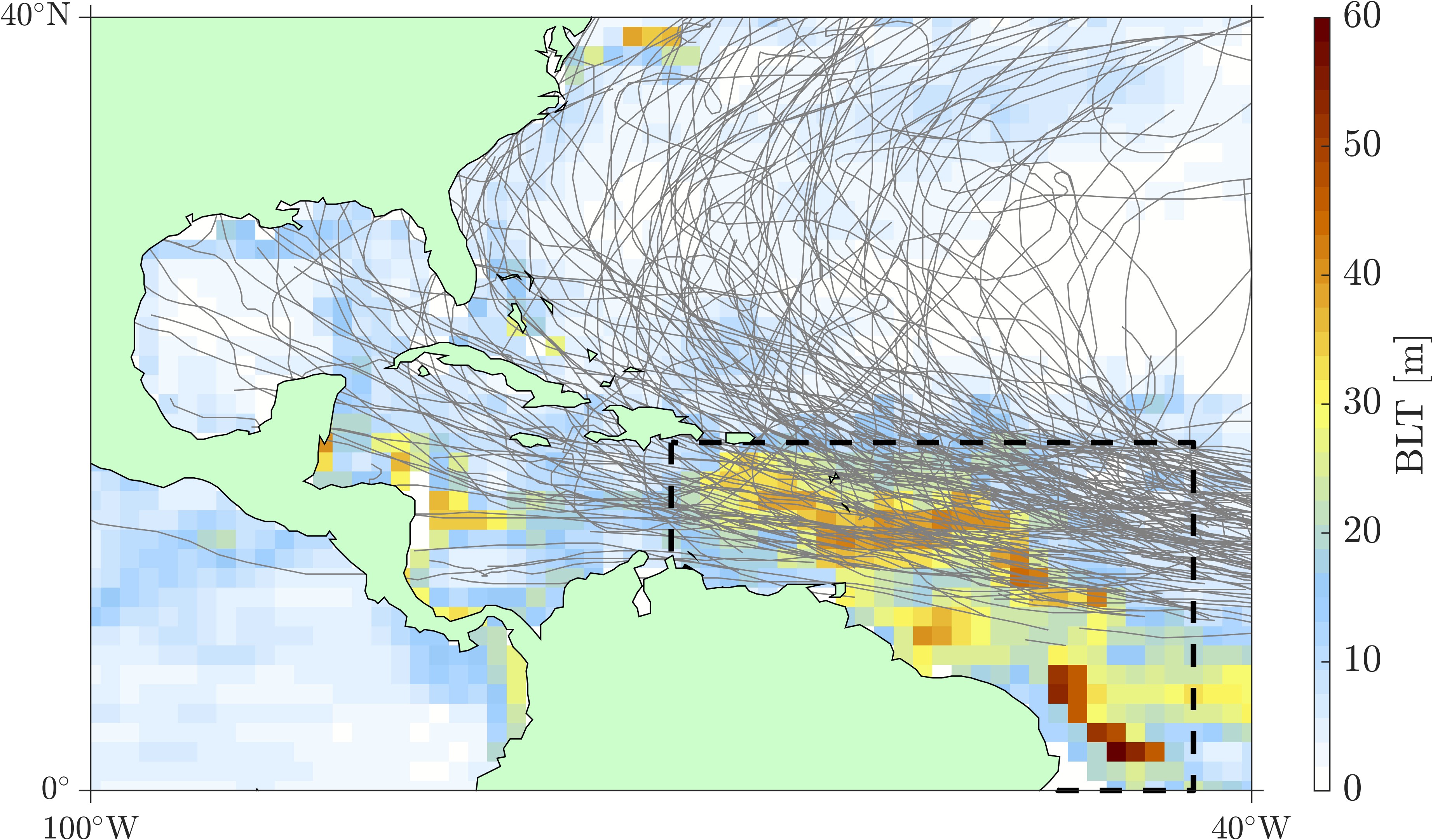}}
        \caption{}
    \end{subfigure}
    \par\bigskip
    \begin{subfigure}{.49\textwidth}
        \includegraphics[width=\textwidth]{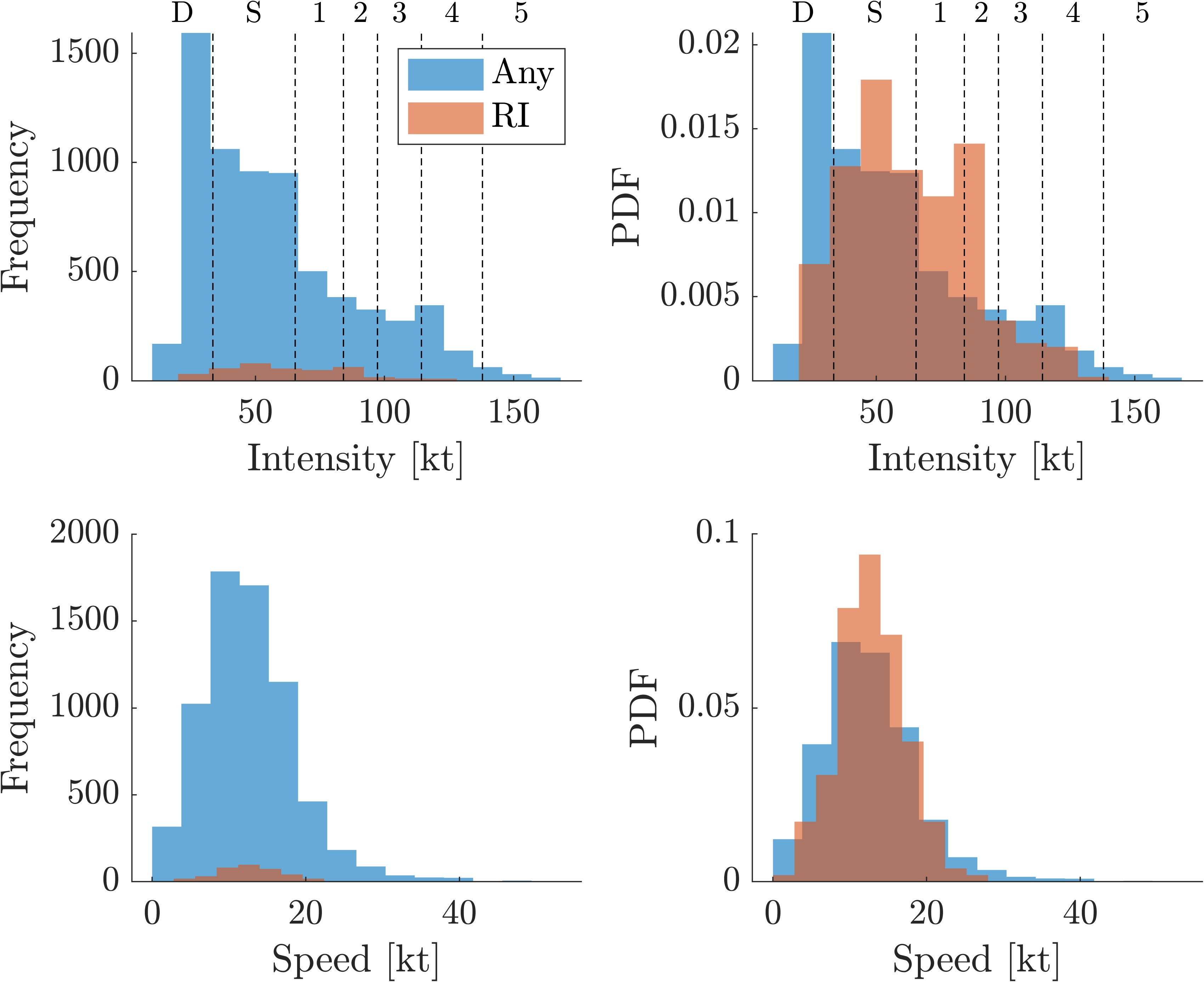}
        \caption{}
    \end{subfigure}
    \begin{subfigure}{.49\textwidth}
        \includegraphics[width=\textwidth]{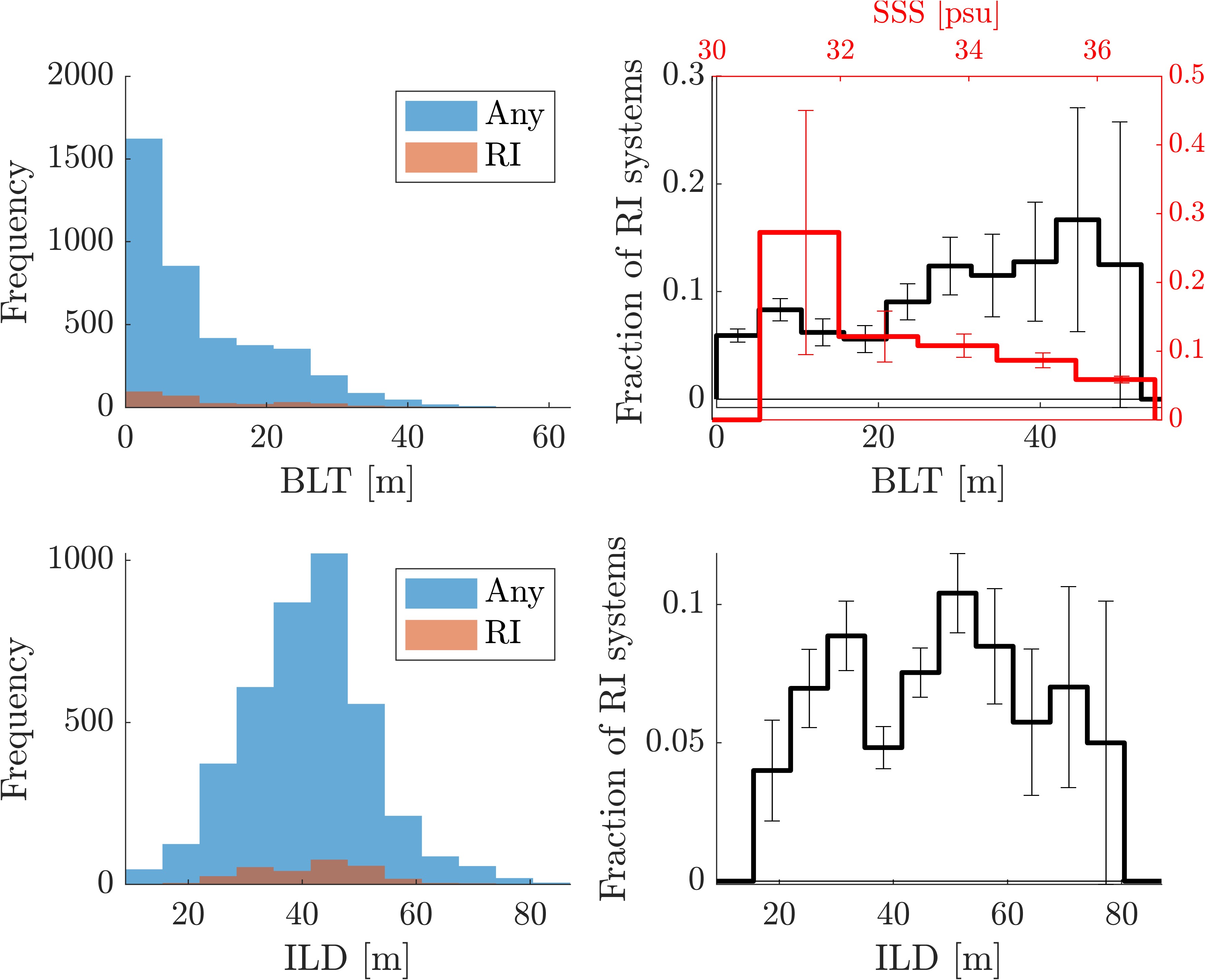}
        \caption{}
    \end{subfigure}
    \par\bigskip
    \begin{subfigure}{.49\textwidth}
        \includegraphics[width=\textwidth]{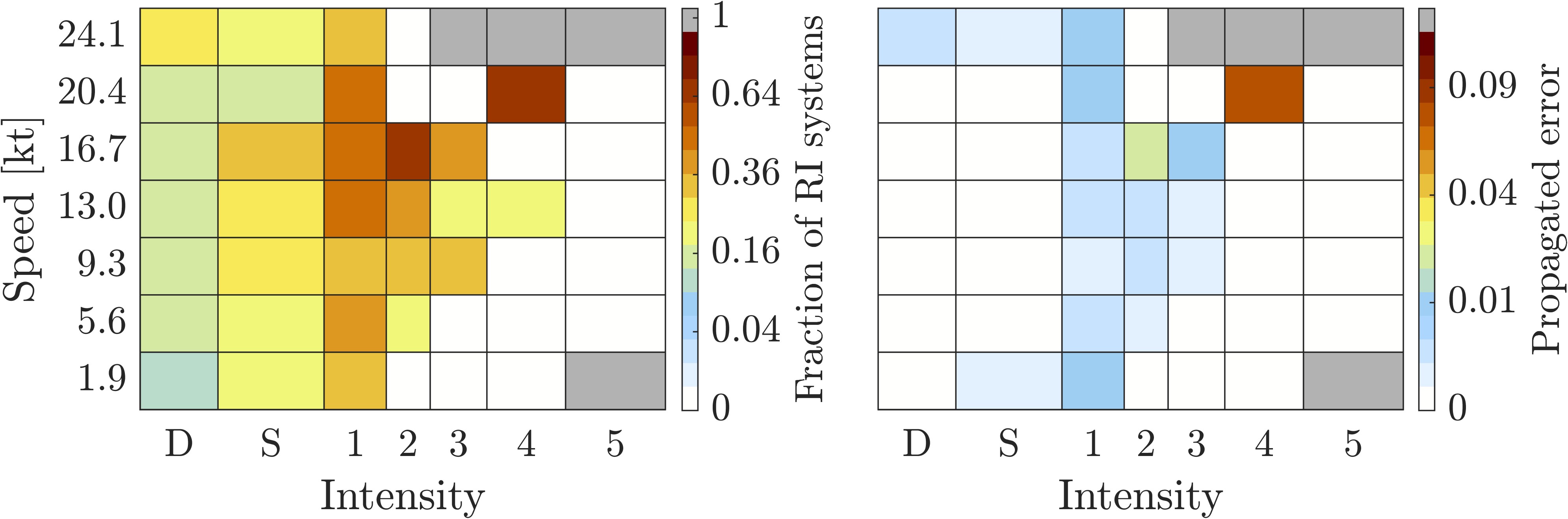}
        \caption{}
    \end{subfigure}
    \begin{subfigure}{.49\textwidth}
        \includegraphics[width=\textwidth]{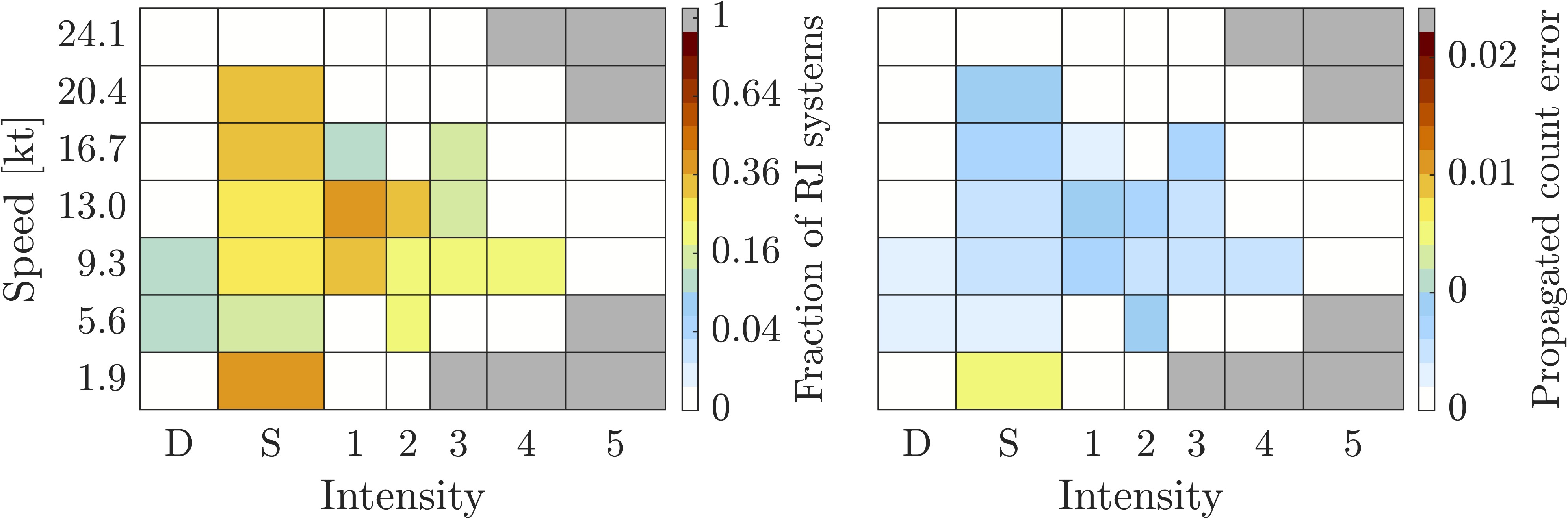}
        \caption{}
    \end{subfigure}
    \caption{\textbf{(a)} Fraction of weather systems exhibiting RI in the Atlantic Ocean (top) and propagated uncertainty (bottom). \textbf{(b)} Overlaid on BL thickness on 15 July 1999, a subset of system tracks considered in this study. \textbf{(c)} Histogram (left) and PDF (right) of along-track intensity (top) and translational speed (bottom). \textbf{(d)} Histogram of along-track BL thickness (top left) and thermocline depth (bottom left) and fraction of systems exhibiting RI for given BL thickness and sea surface salinity (top right) and thermocline depth (bottom right). \textbf{(e)} Fraction of systems exhibiting RI (left) and propagated error (right) as a function of speed and intensity with BL present. \textbf{(f)} Same as \textbf{(e)}, but with BL absent.}
    \label{fig:stat}
\end{figure}

We start by conducting a straightforward calculation of the ratio between the number of weather systems that have undergone RI at some point within a specified $1^\circ \times 1^\circ$ grid box and the total number of weather systems that traverse that grid. For simplicity, this ratio is referred to as \emph{fraction of RI systems}. This is depicted in the upper panel of Fig.\@~\ref{fig:stat}a, while the lower panel illustrates the uncertainty associated with this count, calculated by error propagation. The uncertainty for a count is often approximated as Poisson, $\delta n \approx \sqrt{n}$. For a quantity of interest defined as the fraction $f(n,N)=n/N$, where $n$ is the number of samples belonging to a given class out of a total of $N$ samples, the uncertainty is estimated using a binomial model. In this setting, $n$ is a subset of $N$, so the two counts should not be treated as independent. The resulting uncertainty in $f$ is
\begin{equation}
    \delta f
    =
    \sqrt{\frac{f(1-f)}{N}}
    =
    \sqrt{\frac{n(N-n)}{N^3}},
    \label{eq:error}
\end{equation}
which we display for future reference.  The computation of the fraction of RI systems provides evidence that the Caribbean Sea and the adjacent tropical Atlantic are regions where RI occurs.

In the subsequent analysis within this section and the specialized probabilistic assessment in the following section, we focus on the trajectories of weather systems that have traversed the region outlined by the dashed line in Fig.~\ref{fig:stat}b. This particular area has been selected due to its direct influence from freshwater discharge from the Orinoco and Amazon rivers, making it especially prone to the formation of barrier layers. The figure illustrates the system tracks that fulfill the stated condition, overlaid on the BLT on a randomly chosen day, 15 July 1999. Note the pronounced BL, which exhibits a thickness of up to 60 m. It is systematically organized in a plume-like configuration, originating from the area near the mouth of the Amazon River. Although not all systems advance into the Caribbean Sea, a significant number do so.

By focusing on the weather systems with corresponding tracks depicted in Fig.\@~\ref{fig:stat}b, we calculated histograms and probability density functions (PDFs) for along-track Intensity and Speed, irrespective of whether RI occurred or the timing of such events. These analyses are shown in the top and bottom panels of Fig.\@~\ref{fig:stat}c. The label ``Any'' refers to scenarios regardless of whether RI occurs, while the label ``RI'' denotes the cases where RI occurs. The primary observation is that RI is observed in approximately 5\% of the instances.  This percentage may seem small, but it is important to realize that it does \emph{not} reflect the relative number of systems that experienced RI, which is approximately 40\%, consistent with previous evaluations \citep{Lee-etal-16}. Rather, it reflects the relative times along a system track over the entire collection of tracks that RI is experienced. With this clarification in mind, when focusing on intensity, we grouped it by system category, revealing that RI is detected more frequently in tropical storms (S) and up to category 2 hurricanes (2). With respect to speed, no significant differences are apparent when focusing exclusively on systems experiencing RI. Regardless of the occurrence of RI, the PDF of Speed consistently displays a narrow distribution centered around 12.5\,kt.

Based on our earlier hypotheses, we expect that analyzing along-track BLT and ILD data will be particularly revealing. With this expectation, we have computed histograms for these variables that are depicted in the top and bottom-left panels of Fig.\@~\ref{fig:stat}d. These are presented by categorizing scenarios where RI has occurred (``RI'') and all scenarios regardless of RI occurrence (``Any''). Given the rarity of RI, it is perhaps unsurprising that these histograms highlight the obvious conclusion: weather systems typically sample ocean regions with an ILD ranging from 20 to 60\,m, and a thin barrier layer---a common scenario. More insightful, however, is the ratio of RI systems---the proportion of systems undergoing RI at any given time relative to all systems---in relation to BLT and ILD. These plots suggest a connection between RI events and the presence of a BL. A similar analysis of SSS suggests that RI events tend to occur at low SSS levels and when the barrier layer is thick. This is demonstrated by the proportion of RI systems peaking at $\text{BLT} \approx 50$\,m and SSS of about 32\,psu. This supports earlier comments, indicating that low SSS serves as an indicator of a well-developed BL.  In contrast, the relationship between RI and ILD shows no discernible trend in the proportion of systems that undergo RI as a function of ILD. This contradicts the expectation that a greater thermocline depth, that is, a large ILD, would increase the likelihood of RI by providing more energy for weather systems in the water column. However, we acknowledge that these findings have significant uncertainty, primarily due to limited data availability.

We have performed an additional computation on the fraction of RI systems within the Intensity vs.\ Speed space, both in the presence ($\text{BLT} > 5$ m; cf.\ Section 2.2) of a BL (Fig.\@~\ref{fig:stat}e) and its absence (Fig.\@~\ref{fig:stat}f). In the presence of a BL, a seemingly linear relationship between Speed and Intensity is observed, which aligns with the hypothesis discussed in the Introduction. Our results indicate that for a strong weather system to undergo RI, it must move sufficiently fast to minimize its impact on the BL. In other words, a strong system that remains stationary for an extended period will erode the BL by drawing up cooler water, thereby hindering its chances of intensification. In contrast, for a weak system, breaching the BL becomes more difficult despite moving slowly, preserving its potential for RI. In scenarios where BL is absent or significantly thin ($\text{BLT} < 5$ m), no clear relationship between Intensity and Speed is observed.

In summary, the statistical analysis of this section indicates a comparatively elevated frequency of rapid RI in the Caribbean Sea and the nearby tropical Atlantic, influenced by freshwater from river runoff. It shows that RI occurs primarily in tropical storms and category 2 hurricanes. With a well-developed BL, RI is generally favored, albeit with large uncertainty. However, as systems become more intense, they must also move more quickly to experience RI. Without a BL, there is no clear relationship between strength and translational speed that provides favorable conditions for RI.  Although limited by data availability, the specialized probabilistic analysis in the next section identifies a hidden link between RI and thermocline depth. This connection is not evident in the basic statistical analysis and proves to be more critical than the presence of a BL, contrary to previous expectations.

\subsection{Transition path theory analysis}\label{sec:tpt}

Given that the incidence of RI along a system track is relatively infrequent, comprising approximately 5\% across the entire ensemble of tracks, RI can be conceptualized as a \emph{rare event}. This observation makes it suitable for analysis using advanced probability theory tools designed to manage its complexities. One such tool is \emph{transition path theory} (TPT), which comprises a set of statistics that can be calculated to investigate \emph{productive} pathways, generally showing minimal detouring, between regions of interest in state space \citep{E-VandenEijnden-06, VandenEijnden-06, Metzner-etal-06, Helfmann-etal-20}.  A convenient framework for implementing TPT is a \emph{Markov chain} \citep{Norris-98, Bremaud-99}.  A Markov chain is a stochastic process in which the likelihood of the system being in a specific state at the next time step is solely influenced by the current state and the given transition probabilities.  Transition pathways are computed under the condition that the transition between a given \emph{source} and \emph{target} region is \emph{forthcoming}. Hence, if we choose a target region in an abstract space defined by a state characterized by a particular range of intensities of the weather system, as classified in Table \ref{tab:class} and considered in the preceding statistical analysis, we can study the transitions between this state and the rare state characterized by systems experiencing RI explicitly. The calculation proceeds in two main steps:
\begin{enumerate} 
    \item we construct a Markov chain on states corresponding to discretized \emph{boxes} covering the phase space defined by observations, and 
    \item we define a set of source and target states in this Markov chain and calculate the relevant TPT statistics. 
\end{enumerate}
Here we give a high-level overview of these calculations---refer to Appendices \ref{sec:tpt-app}--\ref{sec:cdf-computation} for further details.

Let us begin by recalling that a Markov chain in discrete time and space is a sequence of random variables $\{X_n\}_{n\in\mathbb Z}$ defined on a space $\mathbb{S} \subseteq \mathbb{N}$ such that
\begin{equation}
    \Pr(X_{n+1} = j) = \sum_{i\in\mathbb S} P_{ij}\Pr(X_n = i),\quad
    P_{ij} := \Pr(X_{n+1} = j \mid X_n = i),
\end{equation}
where $\Pr$ stands for probability. We call the matrix $P = (P_{ij})_{i,j\in\mathbb S}$ a \emph{transition matrix}.  We will assume that there exists a probability vector $\pi = (\pi_i)_{i\in\mathbb S}$, $\sum_{i\in\mathbb S}\pi_i = 1$, called a \emph{stationary distribution}, such that it is invariant, i.e., $\pi = \pi P$, and limiting, i.e., $\pi = \lim_{k\to\infty}vP^k$ for any probability vector $v$. Furthermore, the Markov chain will be assumed to be in stationarity, that is, $\Pr(X_n = i) = \pi_i$.

To construct $P$, we view our observational data, namely, the five records of along-weather-system-track BLT$(t)$, ILD$(t)$, Speed$(t)$, Acceleration$(t)$, and Intensity$(t)$, as series of short-range trajectories, each with a duration of 6 hours. For each weather system, every pair of time-adjacent observations defines the initial and final points for this trajectory, and the totality of these over all weather systems defines the main dataset. We cover the space of this dataset by rectilinear \emph{boxes} in the appropriate dimension with the understanding that each box corresponds to a single state in the Markov chain. Then, $P_{ij}$ is constructed by straightforward membership accounting: 
\begin{equation}
    P_{ij} \approx \frac{\text{number of trajectories in box $i$ that end up in box $j$}}{\text{number of trajectories in box $i$}}.
    \label{eq:Pcount}
\end{equation}

We take special care to define a privileged box that collects all trajectories that happen to take place during an RI event. This naturally leads to the definition of a target $\mathbb{B}$ exactly equal to this RI state. If we choose our partition of the observational data such that along the dimension corresponding to Intensity the boxes have width aligning with weather systems of specific class, then we can choose our source $\mathbb{A}$ to intersect with the set of boxes in a particular class of weather systems. Therefore, \emph{we have a Markov chain with transition probability matrix $P_{ij}$ computed directly from the data, i.e., the time series \textup{BLT($t$)}, \textup{IDL$(t)$}, \textup{Speed($t$)}, \textup{Acceleration($t$)}, and \textup{Intensity($t$)}, as well as definitions for a source $\mathbb{A}$, corresponding to a given weather system class, and a target $\mathbb{B}$, corresponding to RI.} We can now apply TPT to calculate statistics on productive trajectories from $\mathbb{A}$ to $\mathbb{B}$.

We are primarily interested in \emph{transition times}, that is, we would like to construct a ranking of weather systems by class according to which systems have the highest probability of undergoing RI \emph{the most imminently}. One way to build this ranking is by calculating the expected value of a random variable equal to the first time that the target $\mathbb{B}$---the RI state---is visited by the Markov chain in box $i$, conditional on its trajectory being productive.  This TPT statistic, denoted as $t^{i\mathbb{B}}$, was introduced in \citet{Bonner-etal-23}; its computational formula is outlined in Eq.\@~\eqref{eq:trem_comp}. One limitation of $t^{i\mathbb B}$ is that the distribution of the first visit time may have a very long tail---far longer than the lifetime of the weather system---and hence lead to expected values that are unreasonably large. To resolve this limitation, we propose to compute the \emph{distribution of the time $t$ taken by productive trajectories flowing out from $\mathbb A$ to first visit $\mathbb B$}. We denote the associated cumulative distribution function (CDF) by $C^{\mathbb A\mathbb B}(t)$. In this way, $C^{\mathbb A \mathbb B}(t)$ is a measure of the probability that RI occurs (conditionally) in a time not greater than $t$. This is a new TPT statistic that we introduce in this paper. Its computational formula is provided in Eq.\@~\eqref{eq:CAB}, along with a formal definition and a detailed derivation above. 

\begin{figure}[t!]
    \centering
    \includegraphics[width=\textwidth]{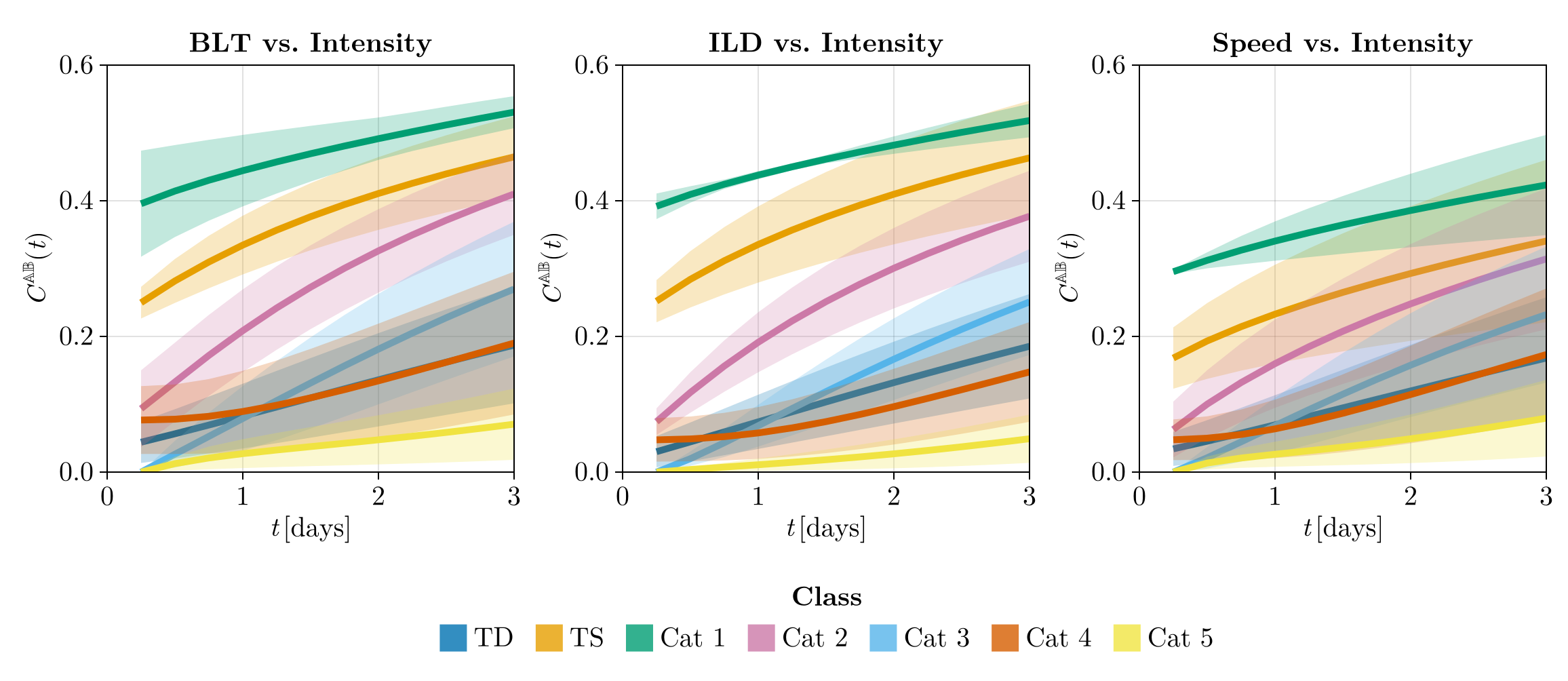}
    \caption{(left panel) CDFs of conditional RI times in a three-day window for each weather system class from a Markov chain generated by trajectory data in (BLT,\,Intensity)-space. The value of the CDF on the vertical axis is the probability that conditional RI occurs on or before the corresponding time on the horizontal axis. (middle panel) As in the left panel, with a Markov chain generated by trajectory data in (ILD,\,Intensity)-space. (right panel) Similar to the left panel, but in (Speed, Intensity)-space. The shaded areas around the curves represent the uncertainty due to error propagation in estimating transition probabilities through counting, as explained in the text.}
    \label{fig:cdf-by-class-weighted}
\end{figure}

Figure \ref{fig:cdf-by-class-weighted} shows $C^{\mathbb{A}\mathbb{B}}(t)$ for each weather system class, derived from Markov chains using three two-dimensional data cross sections: (BLT,\,Intensity), (ILD,\,Intensity), and (Speed,\,Intensity). For the (BLT,\,Intensity) section, we define the source set $\mathbb A$ as the union of the partition boxes that correspond to a given Intensity while covering all BLT box values. This process is done similarly for ILD and Speed. The shading surrounding the curves represents the uncertainty that results from the propagation of the error in estimating the transition matrix of the chain $P$ by counting, as described in Eq.\@~\eqref{eq:Pcount}, as follows. Let $\delta P_{ij}$ represent the propagated error of $P_{ij} = n_{ij}/n_i$, which is calculated according to formula \eqref{eq:error} utilizing $n = n_{ij}$ and $N = n_i$. Subsequently, two matrices $P\pm\delta P$ are derived. Let $C^{\mathbb{A}\mathbb{B}}(t)[P]$ indicate that this TPT statistic is derived from $P$. The corresponding error is then determined as $\delta C^{\mathbb{A}\mathbb{B}}(t)[P] = \frac{1}{2}\smash{\big(C^{\mathbb{A}\mathbb{B}}(t)[P-\delta P] + C^{\mathbb{A}\mathbb{B}}(t)[P+\delta P]\big)}$. To understand Fig.\@~\ref{fig:cdf-by-class-weighted}, it is important to acknowledge that $C^{\mathbb{A}\mathbb{B}}(t)$ is \emph{conditional on RI occurring}.  For example, the left panel of Fig.\@~\ref{fig:cdf-by-class-weighted}, which pertains to the (BLT,\,Intensity) cross section of the data, does not imply that approximately 40\% of Cat 1 systems will undergo RI in a 6-hour window. \emph{Instead, this indicates that for Cat 1 systems committed to experience RI in the future, about 40\% achieve it within a 6-hour period.} Taking this into account, it is feasible to establish a general ranking of RI \emph{imminence} based on the height of the corresponding $C^{\mathbb A \mathbb B}(t)$. This implies that Category 1 systems are the most likely to imminently undergo RI. This finding is consistent with earlier reports based on different assessments \citep{Yan-etal-17}, which found that moderate-strength systems benefit the most from BL presence. In terms of the imminence of undergoing RI, the TPT analysis reveals that Cat 1 systems are followed by TS systems, then Cat 2 systems, and subsequently other classifications of weather systems. This ranking is quite consistent across the (BLT,\,Intensity) and (ILD,\,Intensity) data cross-sections. However, the (Speed,\,Intensity) cross section generally shows lower $C^{\mathbb A \mathbb B}(t)$ values. It is important to note that each data cross section undergoes a distinct Markov chain reduction, resulting in variations in stationary distributions $\pi$; thus, uniformity across $C^{\mathbb A \mathbb B}(t)$ distributions should not be expected.  

\begin{figure}[t!]
    \centering
    \includegraphics[width=\textwidth]{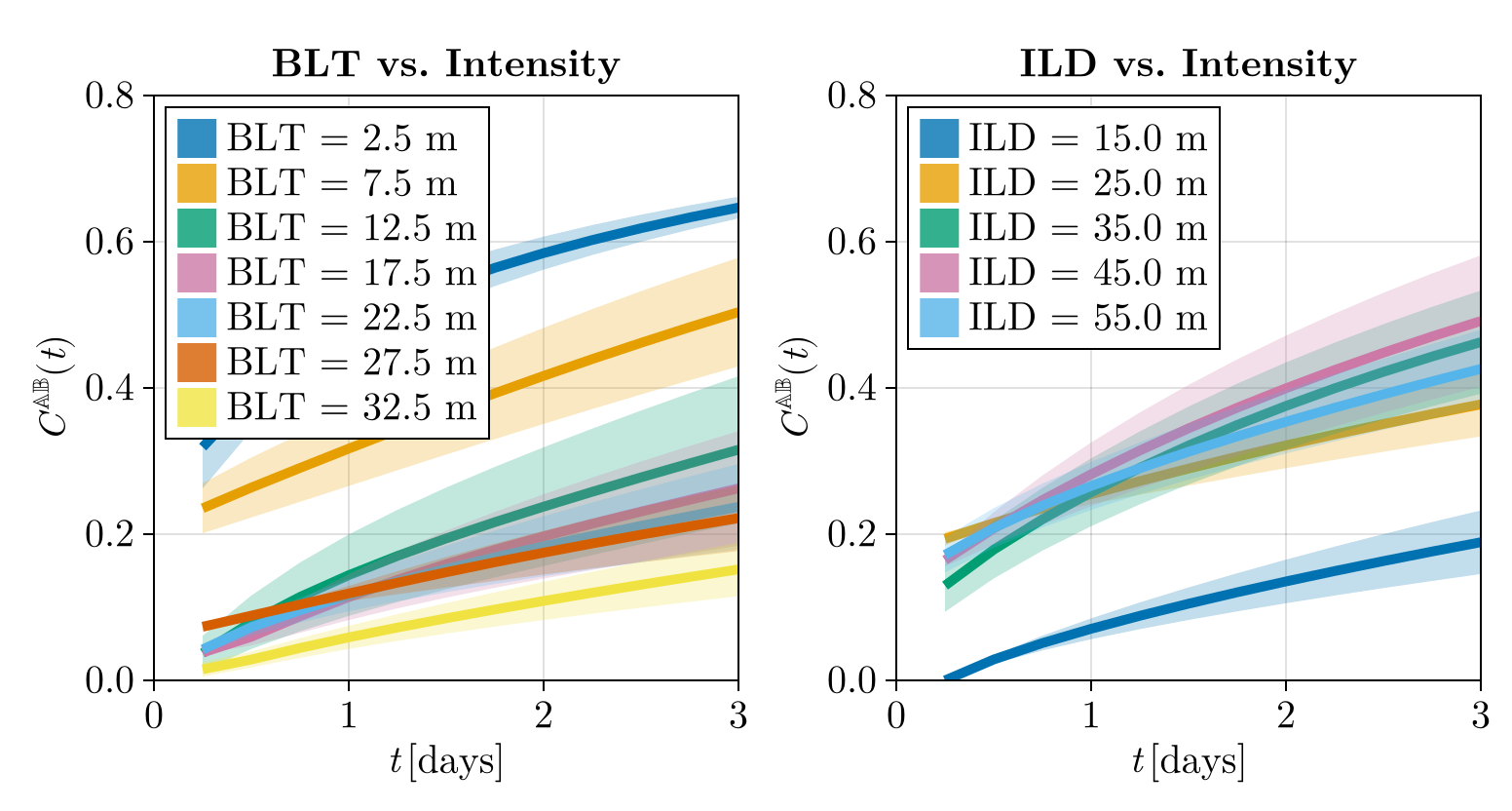}
    \caption{As in Fig.\@~\ref{fig:cdf-by-class-weighted}, but for each BLT (left panel) and ILD (right panel) box values of the Markov chain constructed using trajectory data in (BLT,\,Intensity) and (ILD,\,Intensity)-space, respectively.}
    \label{fig:cdf-by-dens-weighted}
\end{figure}

Although the previous analysis offers new insights into the RI problem, it overlooks the interaction with the density structure of the water column traversed by a weather system. To examine the effects of BLT and ILD, we first consider them independently of the strength of the weather system. Specifically, we calculate $C^{\mathbb{A}\mathbb{B}}(t)$ where $\mathbb{B}$ is the rapid intensification state as before, but now $\mathbb{A}$ is the set of states corresponding to all boxes with a given range of BLT in (BLT,\,Intensity)-space (Fig.~\ref{fig:cdf-by-dens-weighted}, left panel) and similarly for ILD in (ILD,\,Intensity)-space (Fig.~\ref{fig:cdf-by-dens-weighted}, right panel). Large values of $C^{\mathbb{A}\mathbb{B}}(t)$ therefore correspond to imminent RI for weather systems in a particular range of BLT or ILD values averaged over all intensities. Inspection of Fig.\@~\ref{fig:cdf-by-dens-weighted} suggests that RI is more imminent for larger values of ILD, consistent with expectations, but we note it occurs when the BL is thin, contrary to initial expectations and the results of the previous section.

\begin{figure}[t!]
    \centering
    \includegraphics[width=\textwidth]{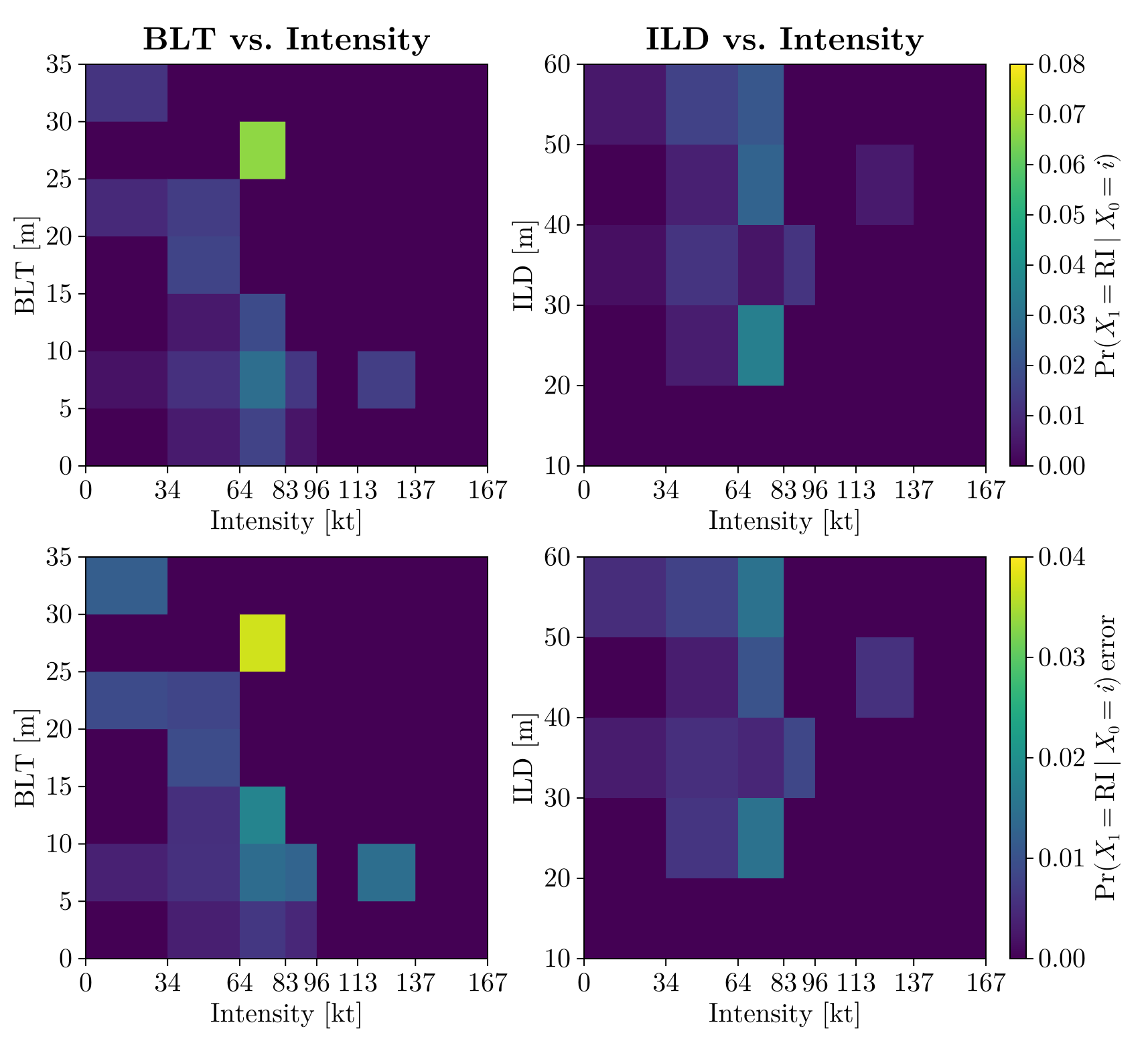}
    \caption{The probability of undergoing RI in the initial step (i.e., after 6 hours) conditional on the Markov chain starting in each box of (BLT, Intensity)-space (top-left panel) and (ILD, Intensity)-space (top-right panel). Uncertainties from error propagation in estimating transition probabilities are shown in the bottom panels.}
    \label{fig:pr-to-ri}
\end{figure}

\begin{figure}[t!]
    \centering
    \includegraphics[width=\textwidth]{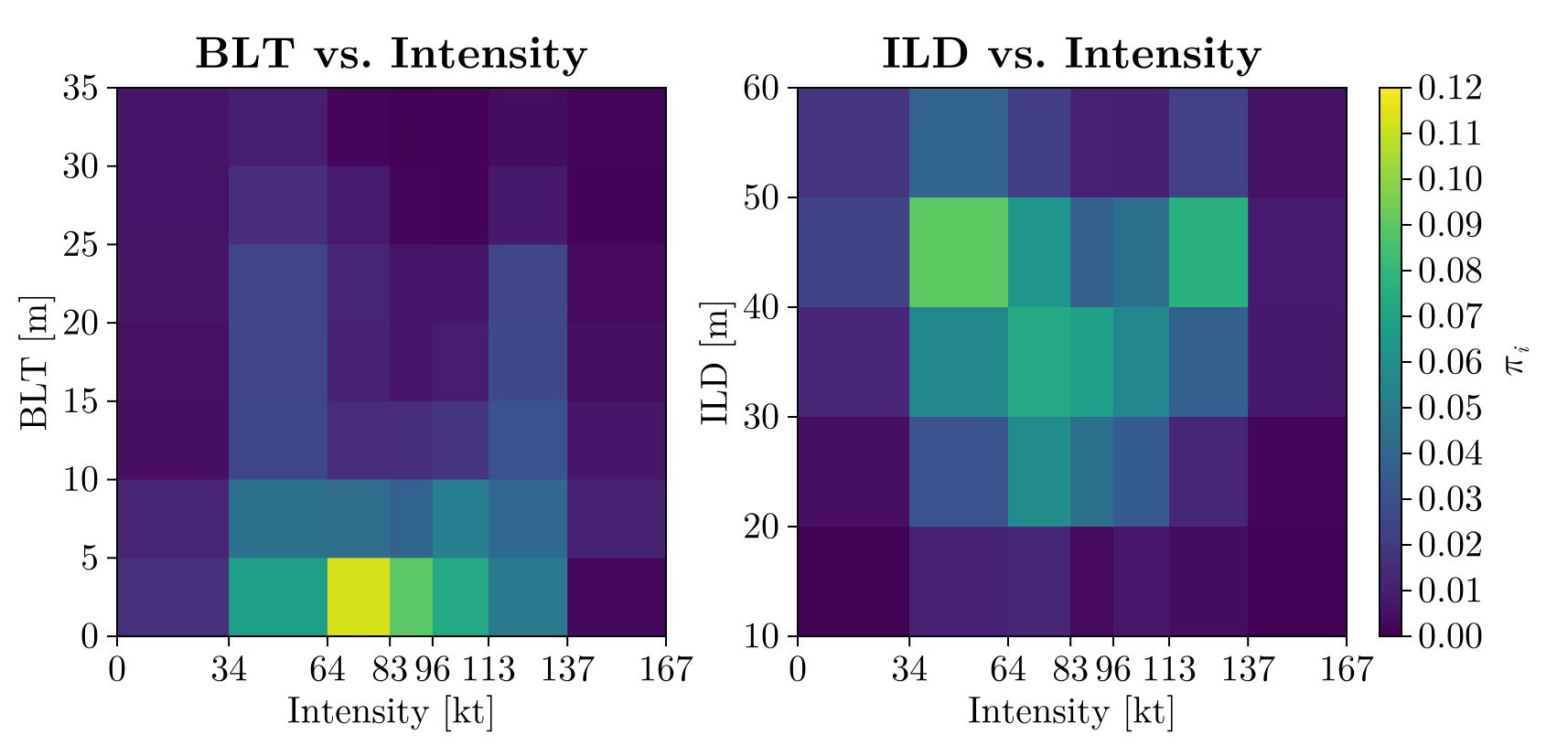}
    \caption{Stationary distribution of the Markov chain in (BLT, Intensity)-space (top-left panel) and (ILD, Intensity)-space (top-right panel).}
    \label{fig:pi}
\end{figure}

To better understand these results, we examine $\Pr(X_1 = \text{RI} \mid X_0 = i)$, the probability of undergoing RI in the first step of the Markov chain, i.e., after 6 hours have elapsed, with an initial state $i$ corresponding to a given box in (BLT,\,Intensity)-space (Fig.\@~\ref{fig:pr-to-ri}, left panel) and (ILD,\,Intensity)-space (Fig.\@~\ref{fig:pr-to-ri}, right panel). Analyzing the left panel of Fig.\@~\ref{fig:pr-to-ri}, $\Pr(X_1 = \text{RI} \mid X_0 = i)$ generally maximizes along rows with low BLT values. Since the stationary distribution ($\pi$) of the Markov chain for the BLT vs.\@~Intensity data cross section (Fig.\@~\ref{fig:pi}, left panel) maximizes when the BL is thin, likely due to the data concentration there, maximization of $\Pr(X_1 = \text{RI} \mid X_0 = i)$ for small BLT suggests high connectivity of these states with the RI state.  The definition of $C^{\mathbb{A}\mathbb{B}}(t)$ implies that $X_1 \notin \mathbb{A}$, hence $C^{\mathbb{A}\mathbb{B}}(t = 6\,\text{h})$ depends directly on $\Pr(X_1 = \text{RI} \mid X_0 = i)$ weighted by $\pi$ over $i \in \mathbb{A}$. Intuitively, short-term imminence will of course be largest for $\mathbb{A}$ with the highest connectivity to the RI state. This appears to explain the tendency of $C^{\mathbb{A}\mathbb{B}}(t)$ to maximize for $\mathbb A$ corresponding to low BLT.  Similar reasoning can explain the tendency of $C^{\mathbb{A}\mathbb{B}}(t)$ to maximize for $\mathbb{A}$ corresponding to high ILD. Note the tendency, though less pronounced, for $\Pr(X_1 = \text{RI} \mid X_0 = i)$, in the right panel of Fig.\@~\ref{fig:pr-to-ri}, to maximize often along rows with high ILD values, where the stationary distribution (Fig.\@~\ref{fig:pi}, right panel) also tends to maximize, likely due to data concentration there. In turn, the results when $\mathbb A = \{\text{Intensity} = \text{const}\}$, as shown in Fig.\@~\ref{fig:cdf-by-class-weighted}, find an analogous interpretation, in this case more evidently, with high connectivity in the second and third columns of the left panel of Fig.\@~\ref{fig:pr-to-ri}. Note that the tendency of $C^{\mathbb{A}\mathbb{B}}(t)$ to maximize for $\mathbb A$ corresponding to Cat 1 weather systems is consistent with $\Pr(X_1 = \text{RI} \mid X_0 = i)$ typically maximizing along the column associated with the Intensity of such systems.  The inferences just made are not exempt from uncertainty, which can be large as shown in the bottom panels of Fig.\@~\ref{fig:pr-to-ri}.

\begin{figure}[t!]
    \centering
    \includegraphics[width=\textwidth]{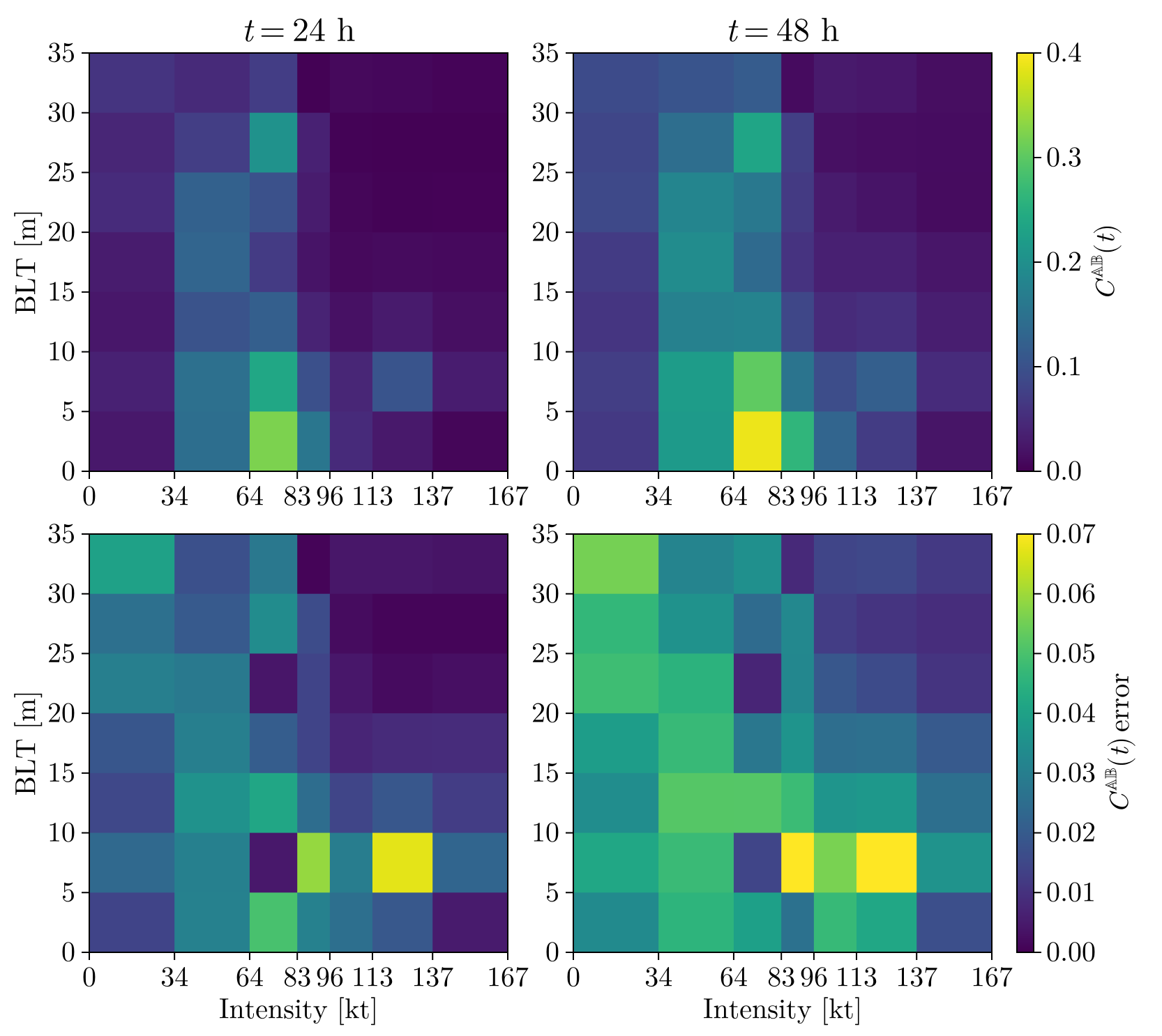}
    \caption{Heatmap showing the conditional RI probability over a 24-hour (top-left panel) and 48-hour (top-right panel) window, derived from a Markov chain on a (BLT, Intensity) cross section. Each box is treated as an independent source $\mathbb{A}$. The bottom panels depict the associated uncertainties, caused by error propagation in estimating transition probabilities through counting.}
    \label{fig:ri-heatmap-blt}
\end{figure}

Computing imminence with source states $\mathbb{A}$ that averaged out a particular variable allowed us to get a general picture of the factors that influence RI. However, this calculation strongly depends on how the initial distribution is chosen and is less applicable to individual weather systems. In order to study the imminence behavior of an arbitrary system more closely, we restrict the size of $\mathbb{A}$ to a \emph{single box} and compute the conditional transition time for RI. Figure\@~\ref{fig:ri-heatmap-blt} presents heatmaps of the $C^{\mathbb{A}\mathbb{B}}(t)$ values derived from a (BLT,\,Intensity) cross section across 24- and 48-hour windows. These are accompanied by the corresponding error metrics that arise from error propagation encountered during the estimation of transition probabilities through counting, as previously explained. In Fig.\@~\ref{fig:ri-heatmap-blt}a, we see that the most imminent transitions to RI tend to occur for low- to medium-intensity systems, say TS to Cat 1--2 systems.  This happens mostly independent of the thickness of the BL; however, there is some preference for states with a thinner BL.  These findings do not fully align---although they do not contradict either---with the anticipated effect of BL presence and RI, which was previously corroborated through the statistical analysis of the preceding section, albeit in an aggregated manner that did not differentiate by weather system class (Fig.\@~\ref{fig:stat}d, top-right panel). The current analysis, which distinguishes by system class, reveals potential deviations from the expected behavior, somewhat consistent with the results shown in Fig.\@~\ref{fig:cdf-by-dens-weighted} obtained by averaging across weather system strength. This is evidenced by the observed peak in RI imminence centered around Cat\,1 systems with a thin BL. 

\begin{figure}[t!]
    \centering
    \includegraphics[width=\textwidth]{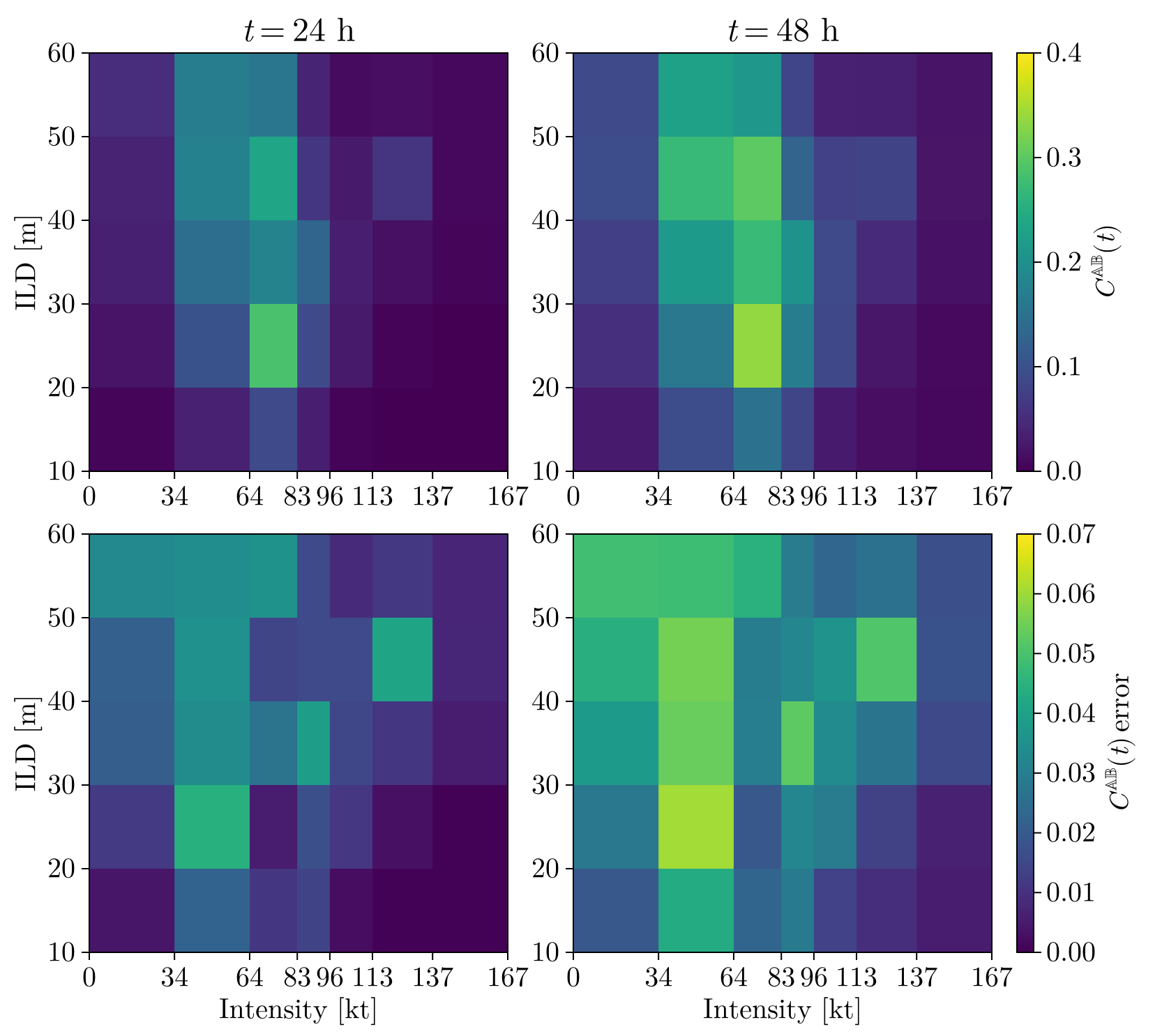}
    \caption{Similar to Fig.\@~\ref{fig:ri-heatmap-blt}, but the Markov chain is constructed on a (ILD, Intensity) cross section.}
    \label{fig:ri-heatmap-ild}
\end{figure}

Figure \ref{fig:ri-heatmap-ild} presents a somewhat more consistent trend through a plot analogous to Fig.\@~\ref{fig:ri-heatmap-blt}, focusing instead on a (ILD,\,Intensity) cross section. The analysis shows that, irrespective of the intensity of the weather system ranging from TS to Cat 1--2, the imminence of RI tends to peak in states characterized by a generally deep IL. Furthermore, the imminence of RI is enhanced in states where the ILD is large compared to those characterized by a thick BLT. This finding is significant because it was not possible to discern any trends using the statistical analysis performed in the previous section (Fig.\@~\ref{fig:stat}d, bottom-right panel), thus demonstrating the significance of the TPT analysis used in the current section.

\begin{figure}[t!]
    \centering
    \includegraphics[width=\textwidth]{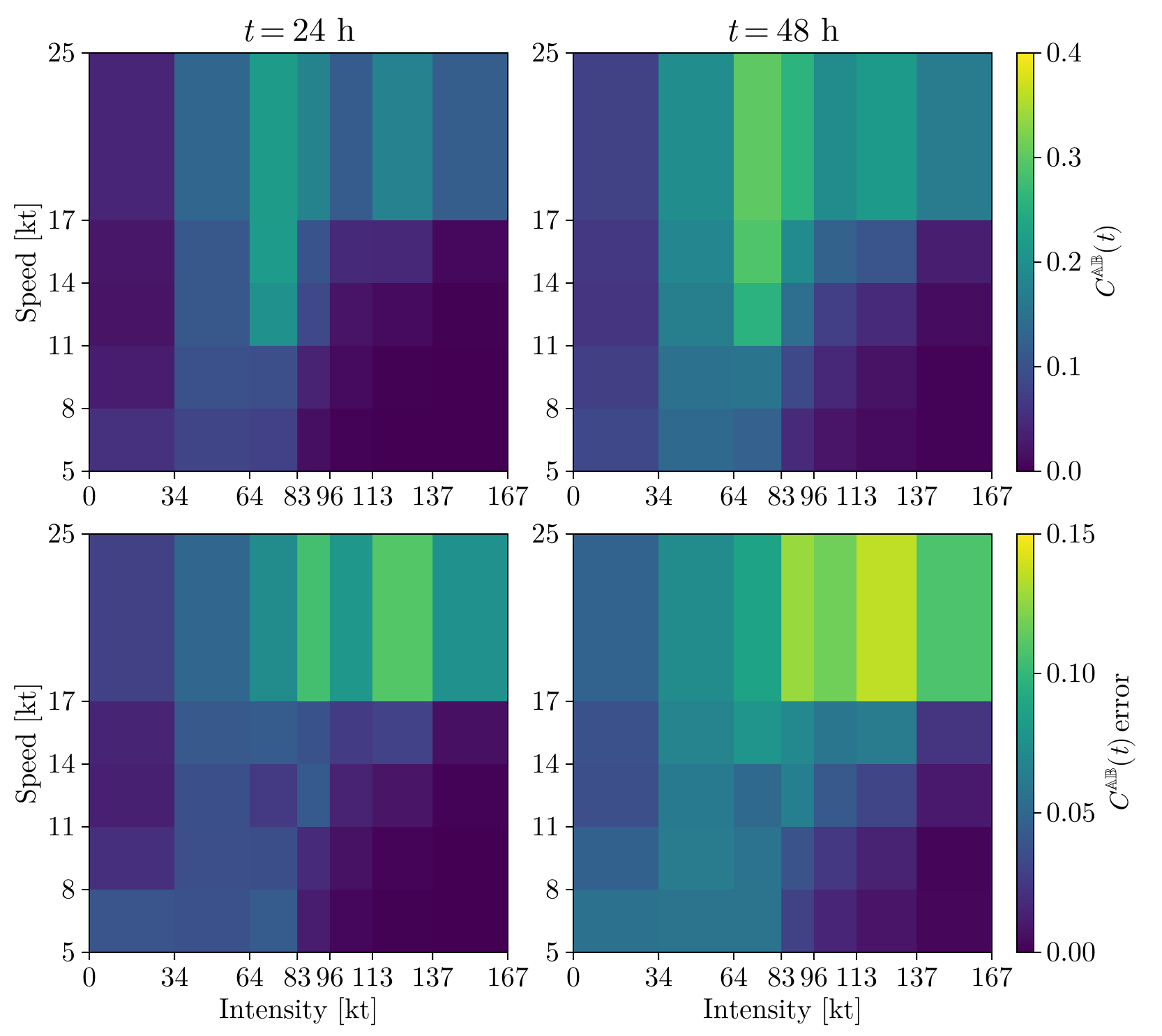}
    \caption{As in Fig.\@~\ref{fig:ri-heatmap-blt}, but for a Markov chain constructed on a (Speed,\,Intensity) cross section.}
    \label{fig:ri-heatmap-speed}
\end{figure}

An additional analysis on how translational speed affects the RI imminence of weather systems was conducted, resulting in the heatmaps shown in Fig.\@~\ref{fig:ri-heatmap-speed}a. These plots, similar to those in Fig.\@~\ref{fig:ri-heatmap-blt}, are based on a (Speed, Intensity) cross section. The heatmaps illustrate that, with the exception of TS systems, a weather system is most imminently subject to RI when it exhibits sufficient translational speed. This phenomenon is particularly pronounced in high-strength weather systems, specifically those classified as Cat 2 to 5. These TPT inference align quite well with the expectation that a stationary strong system has more opportunity to erode the BL and eventually the thermocline below than one moving rapidly through the region. In order for RI to occur, it is essential that a weather system progresses rapidly to inhibit this erosion and the consequent decay, as evidenced by the TPT analysis. This also supports the statistical analysis from the previous section (Figs.\@~\ref{fig:stat}e). The TPT analysis further reveals that the RI imminence is not negligible for slow-moving systems that are weak, specifically for TS to Cat 1 systems.  

\begin{figure}[t!]
    \centering
    \includegraphics[width=\textwidth]{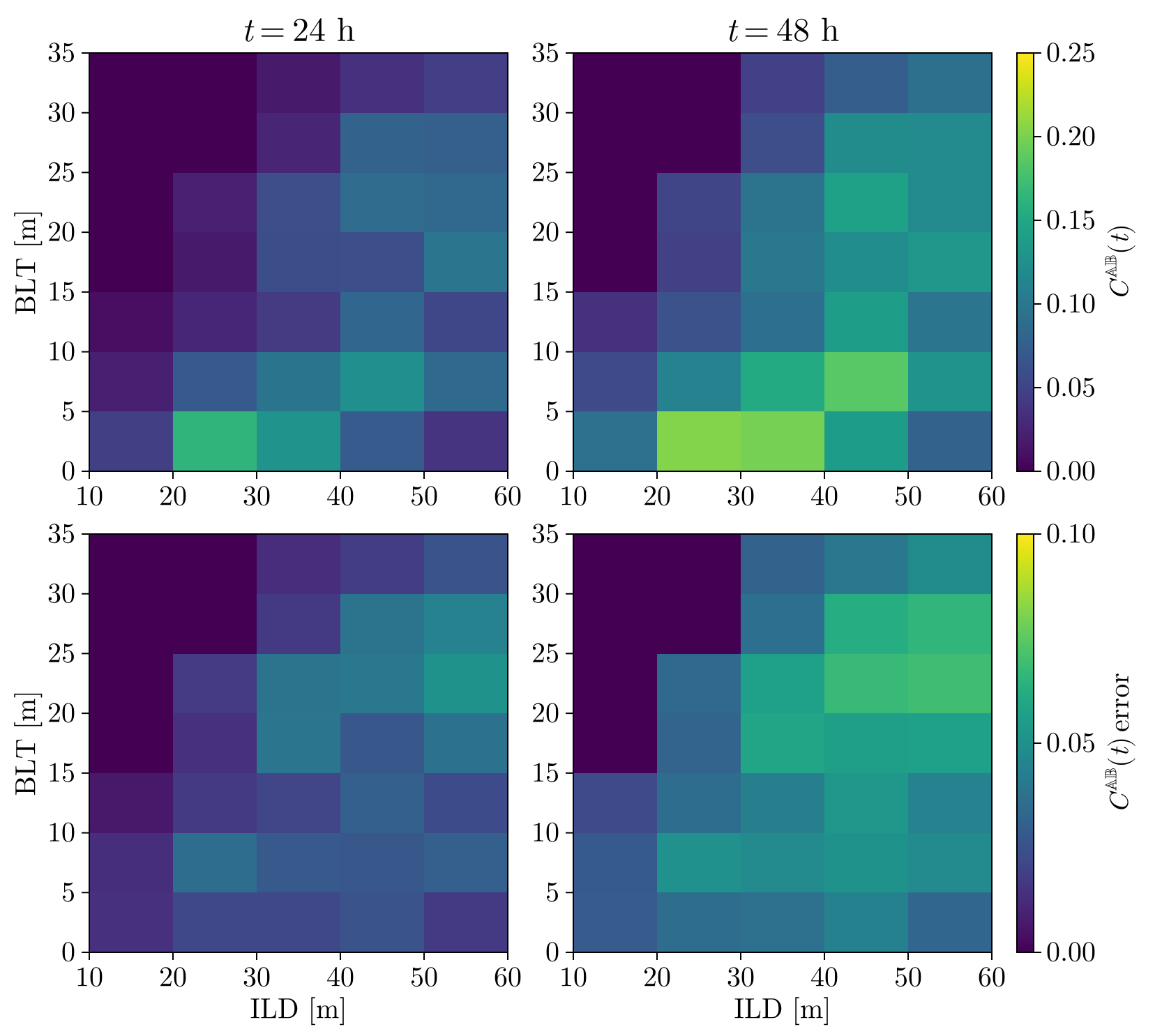}
    \caption{As in Fig.\@~\ref{fig:ri-heatmap-blt}, but for a Markov chain constructed on a (BLT,\,ILD) cross section.}
    \label{fig:ri-heatmap-ild-blt.pdf}
\end{figure}

The results from the final analysis are presented in Fig.\@~\ref{fig:ri-heatmap-ild-blt.pdf}, highlighting RI imminence independent of the strength and speed of weather systems. Specifically, this figure shows heatmaps of $C^{\mathbb{A}\mathbb{B}}(t)$ values derived from a (BLT,\,ILD) cross-section across 24- and 48-hour windows. It is observed that systems of any category, regardless of their speed, generally tend to experience RI when the BL is thick and the thermocline is deep. However, there is a significant probability of imminent RI when the BL is thin and the thermocline is not too deep, contrary to expectations.

In summary, the TPT analysis shows that medium-intensity weather systems, especially Cat 1, have the highest likelihood of RI regardless of the thickness of the BL, thermocline depth, or translational speed. This aligns with assessments reported earlier \citep{Yan-etal-17}. The TPT analysis further highlights a tendency for RI to occur when the thermocline is deep, a result not evident from basic statistical analysis. Surprisingly, a thin barrier layer also maximizes RI imminence. This is explained by a stronger link between deep thermoclines and thin BLs to the RI state in the Markov chain. Fast-moving systems (Cat 2 to 5) are more prone to RI, aligning with expectations that they disturb the water column less. Slow-moving systems (TS to Cat 1) have a lower RI likelihood. Generally, RI happens with a thick BL and deep thermocline, though it can occur with a thin BL, contrary to expectations.

\section{Summary and conclusions}\label{sec:conclusions}

In this paper, we applied two types of analyses on hurricane and reanalyzed hydrographic data to explore the relationship between rapid intensification (RI) and upper ocean density structure, as characterized by the thickness of the barrier layer (BL) and the depth of the thermocline. One analysis involved a straightforward statistical examination of various time series that represent the weather system's strength, translational speed, and rate of change of strength, contrasted with BL thickness and thermocline depth interpolated along system paths. The other analysis was a specialized probabilistic approach based on transition path theory (TPT), which frames RI imminence by reducing the time series viewed as trajectories in different state spaces (cross sections of the data) into Markov chains.  RI imminence is characterized by a newly derived TPT statistic, which represents the probability distribution of the time taken to first visit a target, corresponding to the RI state, from a source, as defined by a specific state in a given data cross section, conditional on the connecting paths being productive, i.e., showing minimal detours.

The direct statistical analysis highlighted a comparatively elevated frequency of RI in the Caribbean Sea and nearby tropical Atlantic, influenced by freshwater from river runoff. It shows that RI occurs primarily in tropical storms and category 2 hurricanes. With a well-developed BL, RI is generally favored, albeit with large uncertainty. However, as systems become more intense, they must also move more quickly to experience RI. Without a BL, there is no clear relationship between strength and translational speed that provides favorable conditions for RI. Although limited by data availability, the TPT analysis identifies a hidden link between RI and thermocline depth. This connection proved to be more critical than the presence of a BL, contrary to previous expectations.

Specifically, the TPT analysis has revealed that regardless of the thickness of the BL, thermocline depth, or translational speed of the weather system, medium-intensity weather systems---dominated by category 1 hurricanes, followed by tropical storms and category 2 hurricanes---have the highest probability of undergoing RI when they are committed to it. For weather systems averaged across strength categories, RI is most likely to occur imminently when the thermocline is deep. Although this result aligns with expectations, it was not discernible through the basic statistical analysis conducted in the previous section. Contrary to expectations, it was found that the likelihood of RI is maximized when the BL is thin. These findings are explained by the stronger direct connection of deep thermocline and thin BL regions to the RI state in the Markov chain. After narrowing the initial distribution to a single box in the chain, we again observed the largest 24- and 48-hour imminence values when the thermocline is deep. Computing RI imminence with initial distributions not spanning multiple states revealed a tendency for RI to occur imminently when the BL is thicker, aligning with expectations. However, category 1 hurricanes showed a strong tendency to imminently undergo RI when the BL is quite thin. The TPT analysis identified translational speed as a significant discriminative parameter. It showed that fast-moving weather systems, particularly the stronger ones (category 2 to 5 hurricanes), are more likely to undergo RI, aligning with expectations that these systems cause minimal disturbance to the water column, thereby reducing the chances of strength decay. The analysis also indicated that slow-moving, weaker systems (tropical storms to category 1 hurricanes) can experience RI, but it is much less likely. Finally, regardless of the system's strength and translational speed, RI typically occurs with a thick BL and deep thermocline. However, RI can often occur even with a thin BL.

Contrary to initial expectations \citep{Balaguru-etal-20}, our results correspond with modeling studies that suggest the BL plays a limited role in RI \citep{Hernandez-etal-16}. Nonetheless, our results are possibly constrained by the scarcity of in-situ observations, which affects the ORAS5 ocean reanalysis data used here. Comparisons of temperature and salinity vertical profiles from ORAS5 with those collected by the Global Ocean Ship-Based Hydrographic Investigations Program (GO-SHIP) \citep{Sloyan-etal-19} show reasonable agreement in the western tropical North Atlantic near the eastern Caribbean. Still, significant discrepancies in surface and subsurface salinities are apparent in the eastern Caribbean. This reveals potential limitations of the reanalysis products in this region and highlights the necessity for more in-situ observations.

\section*{Acknowledgments} 

We sincerely thank Gustavo Jorge Goni for his constant support and commitment to fundamental research. FJBV, GB, and MJO received financial support from the University of Miami's Cooperative Institute for Marine and Atmospheric Studies. HL and SD were supported by NOAA's Atlantic Oceanographic and Meteorological Laboratory (AOML) and the Global Ocean Monitoring and Observing (GOMO) Program.

\section*{Availability availability statement}

The NOAA's NHC HURDAT2 hurricane intensity and speed records are available from \href{https://www.nhc.noaa.gov/data/}{https://\allowbreak www.nhc.noaa.gov/\allowbreak data/}.  The ECMWF's ORAS5 reanalysis data are distributed via \href{https://www.ecmwf.int/en/forecasts/dataset/ocean-reanalysis-system-5}{https://\allowbreak www.ecmwf.int/\allowbreak en/\allowbreak forecasts/\allowbreak dataset/\allowbreak ocean-reanalysis-system-5/}. The GWS Oceanographic Toolbox resides at \href{http://www.teos-10.org/software.htm}{http://\allowbreak www.teos-\allowbreak 10.org/\allowbreak software.htm}. The transition path theory analysis was carried out using the Julia packages \href{https://github.com/70Gage70/UlamMethod.jl}{UlamMethod.jl} and \href{https://github.com/70Gage70/TransitionPathTheory.jl}{TransitionPathTheory.jl} developed by GB with support from NSF under grant OCE2148499.

\appendix
\numberwithin{equation}{section}

\section{Transition Path Theory (TPT)}\label{sec:tpt-app}

\subsection{Reduction of the dynamics into a Markov chain}

We consider data sets consisting of disconnected trajectories $\{x_1(t), x_2(t), \dotsc, x_M(t)\}$ in $\mathcal X \subset \mathbb R^n$.  Each trajectory is sampled at regular time intervals $\Delta t$. We assume that each trajectory is produced by the same underlying \emph{stationary random process}.  We can then think of the existence of an \emph{autonomous} dynamical system $\Phi$, \emph{nondeterministic} in nature, which acts on the phase space $\mathcal X$, representing a measure space equipped with normalized Lebesgue measure $m$, such that $\Phi(x)$ is an $\mathcal X$-valued \emph{random variable} over some implicitly given probability space, with probability measure $\mathbb P$ \citep{Schutte-etal-16}. Let $f \in L^1(\mathcal X)$ be (almost-everywhere) nonnegative and normalized such that $\int_{\mathcal X} f(x)\,m(dx) = 1$. We call such a function a \emph{density}. Let $K(x,y): \mathcal X \times \mathcal X \to \mathbb R^+$, such that $\int_{\mathcal X} K(x,y)\,m(dy) = 1$ for all $x\in\mathcal X$, be the \emph{stochastic kernel} of $\Phi$.  That is, $\mathbb P(\Phi(x)\in\mathcal A) = \int_\mathcal{A} K(x,y)\,m(dy)$, where $\mathcal A$ is a subset of (the $\sigma$-algebra of) $\mathcal X$.  This means, in other words, that $\Phi(x)$ is distributed as $K(x,\cdot)$ or $\Phi(x) \sim K(x,\cdot)$. Then we can define the \emph{Perron--Frobenius operator}, broadly known as a \emph{transfer operator}, $\mathcal P: L^1(\mathcal X) \to L^1(\mathcal X)$ as \citep{Lasota-Mackey-94}
\begin{equation} \label{eq:PFO_def}
    \mathcal P f(y) = \int_{\mathcal X} f(x) K(x,y) \, m(dx).
\end{equation}
This operator describes how an initial density is pushed forward by the underlying stochastic dynamics. We can study the action of $\Phi$ numerically by discretizing $\mathcal P$ using the known trajectory data.

The most widely used discretization scheme is \emph{Ulam's method} \citep{Ulam-60}. Consider a partition $\mathbb X$ of $\mathcal X$ given by the disjoint union of connected boxes $\{B_1, B_2, \dotsc, B_N\}$ and let $\mathbf 1_{\mathbb S}(x)$ be the indicator function on the set $\mathbb S$, which gives 1 when $x\in\mathbb S$ and 0 otherwise. Ulam's method can be interpreted as a Galerkin projection of $\mathcal P$ onto the subspace spanned by $\{\mathbf 1_{B_1}, \mathbf 1_{B_2}, \dotsc, \mathbf 1_{B_N}\}$. By choosing basis functions $\{m(B_i)^{-1} \mathbf 1_{B_i}(x)\}$, we have that the discretization of $\mathcal P$ is an $N$-dimensional linear operator $P$ given by a matrix $(P_{ij}) \in \mathbb R^{N\times N}$ such that $P_{ij}$ is the conditional probability of transitioning from $B_i$ to $B_j$. Since $\mathcal P$ acts on $\{m(B_i)^{-1} \mathbf 1_{B_i}(x)\}$, by Eq.\@~\eqref{eq:PFO_def} we have \citep{Miron-etal-19-JPO, Miron-etal-19-Chaos}
\begin{equation}\label{eq:pij-transfer}
    P_{ij} = \int_{B_j} \mathcal P \frac{\mathbf 1_{B_i}(x)}{m(B_i)} \, m(dx) = \frac{1}{m(B_i)} \int_{B_i} \int_{B_j} K(x,y) \,m(dx)m(dy).
\end{equation}
The matrix $P$ is a row-stochastic \emph{transition matrix} which is the discretized analogue of $K(x,y)$. Note that the factor $m(B_i)^{-1}$ in the choice of basis functions is what ensures that $P$ is (row) stochastic. 

For computational purposes, we approximate Eq.\@~\eqref{eq:pij-transfer} in terms of the trajectory data as \citep{Miron-etal-19-JPO, Miron-etal-19-Chaos}
\begin{equation} \label{eq:pij-boxes}
    P_{ij} \approx \frac{\sum_{m = 1}^M \sum_{t} \mathbf 1_{B_i}(x_m(t)) \mathbf 1_{B_j}(x_m(t + T))}{\sum_{m = 1}^M \sum_{t} \mathbf 1_{B_i}(x_m(t))},
\end{equation}
where $T$ is some multiple of $\Delta t$. The transition probability matrix $P$ defines a \emph{Markov chain} on $N$ boxes representing its \emph{states}. More specifically, the $i$th state is thought of as a delta distribution of (probability) mass located at the center of box $B_i$.

\subsection{The main objects of TPT}

\begin{figure}[t!]
    \centering
    \includegraphics[width=.75\linewidth]{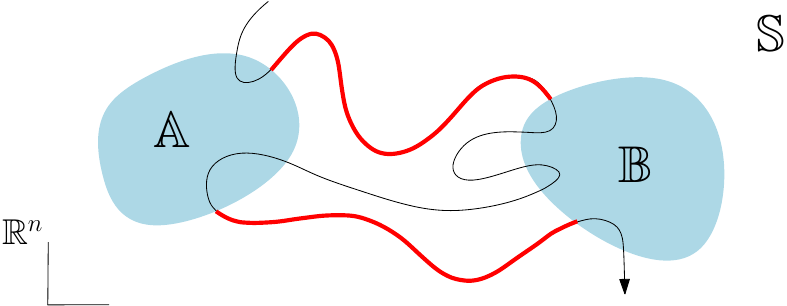}
    \caption{Given a Markov chain taking values on $\mathbb S$, the cartoon shows in red the reactive pieces of a trajectory connecting disjoint target $\mathbb A$ and source $\mathbb B$ subsets of $\mathbb S$.}
    \label{fig:tpt-reactive-def}
\end{figure}

Here we present the central results of TPT; additional details can be found in \citep{E-VandenEijnden-06, VandenEijnden-06, Metzner-etal-06, Helfmann-etal-20}.  To begin, let $\mathbb S = \{1,2,\dotsc,N\}$ represent a finite state space.  Let in addition $X_n$ represent, at discrete time $n$, an $\mathbb S$-valued random variable over an implicitly given probability space, with probability measure given by $\Pr$.  Consider the discrete Markov chain $\{X_n\}_{n \in \mathbb Z}$ with row-stochastic transition probability matrix $P = (P_{ij})_{i,j\in\mathbb S}$. It is assumed that the Markov chain is both ergodic (irreducible) and mixing (aperiodic), and homogeneous in time. It follows that there exists a unique stationary distribution $\pi$ which is invariant, $\pi = \pi P$, and limiting, $\pi = \lim_{k\to\infty} vP^k$ for any probability vector $v$. We take $X_0 \sim \pi$ so that our Markov chain is stationary, that is, we have $X_n \sim \pi P^n = \pi$ \emph{for all} $n \in \mathbb Z$. We define the \emph{first entrance time} to a set $\mathbb D \subset \mathbb S$ as
\begin{equation} \label{eq:first-passage-forward}
    \tau_{\mathbb D}^{+}(n) := \inf\{k \geq 0 : X_{n + k} \in \mathbb D\},
\end{equation}
and the \emph{last exit time} from $\mathbb D$ as
\begin{equation}
    \tau_{\mathbb D}^{-}(n) := \inf\{k \geq 0 : X_{n - k} \in \mathbb D\}.
\end{equation}
The last exit time is a stopping time with respect to the time-reversed process $\{X_{-n}\}_{n \in Z}$, namely, the Markov chain on $\mathbb S$ with transition probability matrix $P^{-} = (P_{ij}^{-})_{i,j\in\mathbb S}$ whose entries are given by
\begin{equation}
    P^{-}_{ij} := \frac{\pi_j}{\pi_i} P_{ji}.
\end{equation}
Let $\mathbb A, \mathbb B$ be two nonintersecting subsets of $\mathbb S$ such that neither is reachable in one step starting from the other. Following the nomenclature used in physical chemistry literature, at time $n$ we say that the process is \emph{forward-reactive} $R^{+}(n)$ (respectively, \emph{backward-reactive}, $R^{-}(n))$ according to the realization of the events 
\begin{equation} \label{eq:reactive-R-definition}
    R^{\pm}(n) := \{\tau_{\mathbb B}^{\pm}(n) < \tau_{\mathbb A}^{\pm}(n)\}.
\end{equation}
Then, the process is \emph{reactive} (referred to as ``productive'' in the more informal terms used in the main text) at time $n$ according to the realization of the event $R(n)$, where
\begin{equation}
    R(n) := \left\{R^{-}(n) \cup R^{+}(n)\right\}.
\end{equation}
In summary, a trajectory is reactive at time $n$ if its most recent visit to $\mathbb A \cup \mathbb B$ was to $\mathbb A$, it is currently outside of $\mathbb A \cup \mathbb B$, and its next visit to $\mathbb A \cup \mathbb B$ will be to $\mathbb B$ (Fig.\@~\ref{fig:tpt-reactive-def}. One can generally think of $\mathbb A$ as a \emph{source} and $\mathbb B$ as a \emph{target} for some process. Associated to the forward and backward reactivities are the \emph{forward} and \emph{backward committors} $q_{i}^{\pm}(n)$ defined for $i \in \mathbb S$ by
\begin{equation} \label{eq:committor-definition}
    q_{i}^{\pm}(n) := \operatorname{Pr}(R^{\pm}(n) \mid X_n = i) .
\end{equation}
We comment briefly on the intuition behind the committors. Note that it is not decidable at time $n$ whether a process is forward reactive at time $n$. If an ensemble of trajectories each have $X_n = i$, then only some fraction of trajectories will hit $\mathbb B$ before $\mathbb A$; this fraction is exactly $q_{i}^+(n)$. States with large values of $q_{i}^+(n)$ tend to be close to $\mathbb B$ in the sense that there is a short path from $i$ to $\mathbb B$ but in general this need not be the case. A similar intuition for $q_{i}^{-}(n)$ holds for the time-reversed Markov chain. The committors are the fundamental quantities of TPT since they contain all of the information about the (infinite) past and future.

One can show \citep{Helfmann-etal-20} that in the case of a homogeneous and stationary Markov chain, the committors are independent of $n$ and satisfy linear matrix equations
\begin{equation}
    q^{+}_i = 
    \begin{cases}
        \sum_{j \in \mathbb S} P_{ij} q_{j}^{+} & \text{if }i \notin \mathbb A \cup \mathbb B,\\ 
        0 & \text{if }i \in \mathbb A,\\ 
        1 & \text{if }i \in \mathbb B,
    \end{cases}
    \quad \text{and} \quad   
    q^{-}_i = 
    \begin{cases}
        \sum_{j \in \mathbb S} P_{ij}^{-} q_{j}^{-} & \text{if }i \notin \mathbb A \cup \mathbb B,\\ 
        1 & \text{if }i \in \mathbb A,\\ 
        0 & \text{if }i \in \mathbb B.
    \end{cases}
\end{equation}
Using the committors, a number of statistics can be computed for reactive trajectories. First, we have the \emph{reactive density}
\begin{equation} \label{eq:muAB}
    \mu^{\mathbb{AB}}_{i}(n) := \operatorname{Pr}(X_n = i,\, R(n)) = q_{i}^{-} \pi_i q_{i}^+, \quad i \notin \mathbb A \cup \mathbb B.
\end{equation}
States with large reactive densities relative to their neighbors are interpreted as bottlenecks for reactive trajectories. We also define the \emph{reactive current}
\begin{equation}
    f^{\mathbb{AB}}_{i j}(n) := \operatorname{Pr}(X_n = i,\, R^{-}(n), X_{n + 1} = j,\, R^{+}(n + 1)) = q_{i}^{-} \pi_i P_{ij} q_{j}^+, \quad i, j \in \mathbb S,
\end{equation}
as well as the \emph{effective reactive current}
\begin{equation} \label{eq:effective-reactive-current}
    f_{i j}^{+} := \max\{f^{\mathbb{AB}}_{i j} - f^{\mathbb{AB}}_{j i}, 0\}.
\end{equation}
The effective reactive current is a $\mathbb B$-facing gradient of the reactive density; it identifies pairs of states with a large net flow of probability. Finally, we have the \emph{transition time} $t^{\mathbb{AB}}$. In its original definition \citep{VandenEijnden-06}, $t^{\mathbb{AB}}$ is given by the limiting ratio of the time spent during reactive transitions from $\mathbb A$ to $\mathbb B$ to the rate of reactive transitions leaving $\mathbb A$. In  \citet{Helfmann-20}, the following expression is provided in the discrete case
\begin{equation} \label{eq:tAB-vanE}
    t^{\mathbb{AB}} := \frac{\operatorname{Pr}(R(n))}{\operatorname{Pr}(R^+(n + 1), X_n \in \mathbb A)} = \frac{\sum_{i \in \mathbb S} \mu_{i }^{\mathbb{AB}}}{\sum_{i \in \mathbb A, j \in \mathbb S} f_{i j}^{\mathbb{AB}}}.
\end{equation}
In \citet{Bonner-etal-23} we introduced a generalization of the transition time which allows one to express Eq.\@~\eqref{eq:tAB-vanE} as a straightforward expectation, which we describe next.

\subsection{Remaining transition time}

The aim of this generalization introduced in \citet{Bonner-etal-23} is to provide local information, that is, information about reactive trajectories at a particular state that have already left the source $\mathbb A$. Let
\begin{equation}
    \mathbb C^+ = \left\{i \notin \mathbb B :  \sum_{\ell \in \mathbb S }P_{i \ell} q_{\ell}^+ > 0 \right\}.
\end{equation}
We define the \emph{remaining transition time} $t^{i\mathbb B}$ for all $n \in \mathbb Z$ as
\begin{equation} \label{eq:trem_def}
    t^{i\mathbb B} 
    :=
    \begin{cases}
        \operatorname{Ex}\left[\tau_{\mathbb B}^+(n + 1) \mid X_{n} = i,\ R^{+}(n + 1)\right] & \text{if }i \notin \mathbb B,\\
        0 & \text{if }i \in \mathbb B.
    \end{cases}
\end{equation}
A similar formula is referred to as the lead time in \citet{Finkel-etal-21}. In \citet{Bonner-etal-23} it is shown that Eq.\@~\eqref{eq:trem_def} satisfies the following set of linear equations:
\begin{equation} 
    t^{i\mathbb B} = 
    \begin{cases}
        1 + \sum_{j \in \mathbb C^+ }\frac{P_{ij} q_{j}^+  }{\sum_{\ell \in \mathbb S }P_{i \ell} q_{\ell}^+} t^{jB} & \text{if }i \in \mathbb C^+,\\
        0 & \text{if }i \in \mathbb B.
    \end{cases}
    \label{eq:trem_comp}
\end{equation}
When $\mathbb A$ contains only one state, we also have that
\begin{equation}
    t^{i\mathbb B} \big|_{i = \mathbb A} = t^{\mathbb{AB}} + 1,
\end{equation}
where $t^{\mathbb{AB}}$ is defined in Eq.\@~\eqref{eq:tAB-vanE}.

\subsection{Open systems and connectivity}

In many problems, the trajectory data are given on an open domain, and further processing is required to obtain a suitable Markov chain. We follow \citet{Miron-etal-21-Chaos} and subsequent works, and introduce a \emph{two-way nirvana state}, $\omega$, to create a closed system. Suppose that all trajectory data are contained inside a domain $\mathbb Y \subset \mathbb X$. We partition $\mathbb Y$ as $\mathbb Y = \mathbb Y^O \cup \omega$ such that $\partial \mathbb Y \subset \omega$ and $|\omega| \ll |\mathbb Y^O|$. We then construct a covering by $N$ boxes of $\mathbb Y^O$ with one additional box appended corresponding to the whole of $\omega$. Applying Eq.\@~\eqref{eq:pij-boxes}, we obtain a row-stochastic transition matrix of the form
\begin{equation}
    P = 
    \begin{pmatrix}
        P^{O \to O} & P^{O \to \omega} \\ 
        P^{\omega \to O} & 0 
    \end{pmatrix},
    \label{eq:Pclose}
\end{equation}
where $P^{O \to O}$ is $N \times N$, $P^{O \to \omega}$ is $N \times 1$ and $P^{\omega \to O}$ is $1\times N$. Note that trajectories which begin and end in the nirvana state are ignored. In general, we are only interested in reactive trajectories which do not visit this extra nirvana state. This requirement is equivalent to making the replacements $\mathbb A \to \mathbb A \cup \omega$ and $\mathbb B \to \mathbb B \cup \omega$ in the computation of $q^+$ and $q^-$, respectively. One can show \citep{Miron-etal-21-Chaos} that this is also equivalent to leaving $\mathbb A$ and $\mathbb B$ unchanged, but replacing $P$ with the row-substochastic matrix $P^{O \to O}$ and $\pi$ by restriction of the stationary distribution of Eq.\@~\eqref{eq:Pclose} to $O$. 

Depending on the shape of the data, $P^{O \to O}$ may not an irreducible, aperiodic matrix. To remedy this, we can apply Tarjan's algorithm \citep{Tarjan-72} to extract the largest strongly connected component of $P^{O \to O}$. Then, $P^{O \to O}$ is modified to remove all other states including contributions from trajectories which pass through the removed states. The final result is that we have an irreducible aperiodic transition matrix which avoids the nirvana state and is suitable for use in the formulation of TPT above.

\section{Application of TPT to the RI problem}\label{sec:tpt-prob}

Consider time series BLT($t$) (barrier layer thickness), ILD($t$) (thermocline depth), Intensity($t$) (maximum sustained wind), and Speed($t$) (of translation), all measured at intervals $\Delta t = 6$\,h along the track of each weather system passing through the region indicated in Fig.\@~\ref{fig:stat}b, as explained in the main text. Let ws be a generic weather system and consider the observation times $t$ of that system to be $t_i^\text{ws} = t_0^\text{ws} + i\Delta t$ for $i \in \{0, 1, \dotsc, N_\text{ws}\}$, where $t_0^\text{ws}$ is the initial time and $N_\text{ws}+1$ is the total number of observations. Our data defines a trajectory in the four-dimensional space composed of the time-ordered collection of points $\{\mathbf x_1^\text{ws}, \mathbf x_2^\text{ws}, \dotsc, \mathbf x_{N_\text{ws}}^\text{ws}\}$ where
\begin{equation}
    \mathbf x_i^\text{ws} = \big(\text{BLT}^\text{ws}_i, \text{IDL}^\text{ws}_i, \text{Intensity}^\text{ws}_i, \text{Speed}^\text{ws}_i\big)
\end{equation}
with $\text{BLT}^\text{ws}_i :=  \text{BLT}^\text{ws}(t_i^\text{ws})$ and similarly for the other variables. We note that the following analysis applies equally well to lower-dimensional subsets of these data, e.g., $\mathbf{x}_i^\text{ws} = (\text{BLT}^\text{ws}_i, \text{Intensity}^\text{ws}_i)$, with minimal modifications. In order to study RI, for each ws we construct a new set of trajectories 
\begin{equation}
    \mathsf R^\text{ws} = \big\{(\mathbf x_1^\text{ws}, r_1^\text{ws}), (\mathbf x_2^\text{ws}, r_2^\text{ws}), \dotsc, (\mathbf x_{N_\text{ws}}^\text{ws}, r_{N_\text{ws}}^\text{ws})\big\},
\end{equation}
where the additional element $r_i^\text{ws}$ is computed as follows. Let $\mathfrak A$ denote the threshold minimum Acceleration---average daily increase in Intensity---necessary for a weather system to be classified as experiencing RI (rapid intensification). Then, 
\begin{equation}
    r_i^\text{ws}
    =
    \begin{cases}
        1 & \text{if Acceleration}^\text{ws}_i > \mathfrak A,\\
        0 & \text{otherwise}.
    \end{cases}
\end{equation}
Intuitively, $r_i^\text{ws}$ is a flag that indicates when observation $\mathbf x_i^\text{ws}$ occurs during a period of RI. Each $\mathsf R^\text{ws}$ can be decomposed into a set of one-step trajectories; for each $i \in \{0, 1,\dotsc, N_\text{ws}\}$ we have the trajectory with initial and final points $(\mathbf x_i^\text{ws}, r_i^\text{ws}) \to (\mathbf x_{i + 1}^\text{ws}, r_{i + 1}^\text{ws})$. For future reference, let $\mathsf R_0$ be the ordered set of all initial points $(\mathbf x, r)_0$ of all one-step trajectories over all $\mathsf R^\text{ws}$. Let $\mathsf R_{\Delta t}$ be the ordered set of all the corresponding final points $(\mathbf{x}, r)_{\Delta t}$.

We can now apply Ulam's method of trajectory discretization. First, let $\mathcal A = \mathcal B \times \{0, 1\}$ where $\mathcal B \subset \mathbb R^4$ is the minimal volume hyperrectangle such that each member of $\mathsf R_0 \cup \mathsf R_{\Delta t}$ is contained in $\mathcal A$.\footnote{This hyperrectangle is unique. For a given set of data in $\mathbb{R}$, there is a unique closed interval of $\mathbb{R}$ of minimal length containing the data whose endpoints are equal to the extrema of the dataset. This argument generalizes to higher dimensions.} Let $\epsilon \in (0, 1]$ and let $\mathcal B_\epsilon \subset \mathcal B$ be the box obtained by uniform scaling of $\mathcal{B}$ by $\epsilon$. Defining $\omega := \mathcal{B}\setminus\mathcal{B}_\epsilon$, we then have $\mathcal A = \mathcal B_\epsilon \cup \omega \times \{0, 1\}$, where $\omega$ represents the nirvana state. We tessellate $\mathcal B_\epsilon$ using a rectilinear grid so that $\mathcal B_\epsilon = \bigcup_{i = 1}^N B_i$ such that $B_i \cap B_j = \varnothing$ for all $i \neq j$. In particular, the tesselation in the dimension corresponding to Intensity is constructed such that the grid points fall within weather system class boundaries as defined in Table~\ref{tab:class}. Our final partition is therefore $\mathcal A = \bigcup_{i = 1}^N B_i  \cup \omega \times \{0, 1\}$. We assign an index to each point $(\mathbf x, r) \in \mathsf R_0 \cup \mathsf R_{\Delta t}$, according to its membership via
\begin{equation}
    I[(\mathbf x, r)] 
    = 
    \begin{cases}
    i & \text{if } r = 0,\, \mathbf x \in B_i,\, i = 1,2,\dotsc,N,\\
    N + 1 & \text{if } r = 0,\, \mathbf x \in \omega,\\
    N + 2 & \text{if } r = 1.
    \end{cases}
\end{equation}
Now we can construct a row-stochastic transition probability matrix $P\in\mathbb R^{(N+2)\times (N+2)}$ whose entries are given by
\begin{equation} \label{eq:transition-matrix-indicators-def}
    P_{ij} = \frac{\sum_\ell \mathbf 1_i(I[(\mathbf x_\ell, r_\ell)_0)]) \mathbf 1_j(I[(\mathbf x_\ell, r_\ell)_{\Delta t})])}{\sum_\ell \mathbf 1_i(I[(\mathbf x_\ell, r_\ell)_0)])}.
\end{equation}

Finally, we define the discrete-time Markov chain $\{X_n\}_{n \geq 0}$ in the set $\{1, 2, \dots, N + 2\}$ with transition probabilities $P(X_{n + 1} = j \mid X_{n} = i) = P_{ij}$. By construction, this is an ergodic Markov chain with a unique stationary distribution $\pi$ that satisfies $\pi = \pi P$ and hence the standard results of stationary TPT apply. We associate the states indexed by $N + 1$ and $N + 2$ with $\omega$ and $\rho$, where the latter refers to the RI state. To apply TPT, a natural choice for the target is $\mathbb B = \{\rho\}$. For any source ($\mathbb A$) under consideration, we append $\omega$ to $\mathbb A$ (resp., $\mathbb B$) when we compute the forward (resp., backward) committor, thereby restricting our attention to trajectories that avoid nirvana.

\section{Computation of $C^{\mathbb{A} \mathbb{B}}(t)$}
\label{sec:cdf-computation}

In order to study transitions from $\mathbb{A}$ to $\mathbb{B}$ more precisely, we compute the \emph{distribution} of a conditional first passage time measuring the time taken for trajectories initialized inside $\mathbb{A}$ to reach $\mathbb{B}$ in a reactive sense. Let $\mathbb{A}^\star = \mathbb{A}\setminus (\mathbb{A}\cap\mathbb{B})$ represent the set of states that are in $\mathbb{A}$ but not part of the set that is avoided (including nirvana)---note that this set is nonempty by the construction of $\mathbb{A}$ and $\mathbb{B}$. Similarly, let $\mathbb{B}^\star = \mathbb{B}\setminus (\mathbb{A}\cap\mathbb{B})$. We compute the following generating function
\begin{align} 
    \mathcal{T}(n; \lambda) &= \operatorname{Ex}[\lambda^{\tau_{\mathbb{B}}^{+}(n)} \mid R^+(n + 1), X_n \sim \pi_{\mathbb{A}^\star}] \label{eq:T-gf-definition} \\
    &= \sum_{\ell = 0}^{\infty} \operatorname{Pr}(\tau_{\mathbb{B}}^+(n) = \ell \Delta t \mid R^+(n + 1), X_n \sim \pi_{\mathbb{A}^\star}) \lambda^\ell, \\ 
    &=: \sum_{\ell = 0}^{\infty} C^{\mathbb{A} \mathbb{B}}(\ell \Delta t) \lambda^\ell,
\end{align}
where $\tau_{\mathbb{D}}^{+}(n)$ is defined in Eq.~\eqref{eq:first-passage-forward}, $R^+(n)$ is defined in Eq.\@~\eqref{eq:reactive-R-definition}, and $X_n \sim \pi_{\mathbb{A}^\star}$ means
\begin{equation} \label{eq:pi-Astar-definition}
    \operatorname{Pr}(X_n  = i \mid X_n \sim \pi_{\mathbb{A}^\star}) 
    = 
    \begin{cases}
        \frac{\pi_{i}}{\sum_{j \in \mathbb{A}^\star} \pi_j} & \text{if } i \in \mathbb{A}^\star \\ 
        0 & \text{otherwise},
    \end{cases}
\end{equation}
where $\pi$ is the stationary distribution of the Markov chain. Before proceeding with Eq.\@~\eqref{eq:T-gf-definition}, we define two conditional probabilities, first
\begin{align}
    P_{ij}^{+} &:= \operatorname{Pr}(X_{n + 1} = j \mid R^+(n + 1), X_n = i) \\
     &= \frac{\operatorname{Pr}(X_{n} = j \mid X_{n} = i) \operatorname{Pr}(R^+(n + 1) \mid X_n = i, X_{n + 1} = j)}{\operatorname{Pr}(R^+(n + 1) \mid X_{n} = i)} \\ 
    & = \frac{P_{ij} q_{j}^+}{\sum_{i \in \mathbb{S}}P_{ij} q_{j}^+}, 
    \label{eq:Pijplus-definition}
\end{align}
where we have used Bayes' rule, the Markov property, and Eq.\@~\eqref{eq:committor-definition}. Similarly, we define
\begin{align}
    Q_{i}^+ &:= \operatorname{Pr}(X_{n} = i \mid R^+(n + 1), X_n \sim \pi_{\mathbb{A}^\star}) \\
    &= \frac{\pi_{\mathbb{A}^\star}(i) \sum_{\ell \in \mathbb{S}}P_{i\ell} q_{\ell}^+}{\sum_{m \in \mathbb{A}^\star}\pi_{\mathbb{A}^\star}(m) \sum_{\ell \in \mathbb{S}}P_{m\ell} q_{\ell}^+}. 
    \label{eq:Qj-definition}
\end{align}
Now, Eq.\@~\eqref{eq:T-gf-definition} implies by conditioning on the value of $X_n$ and then conditioning on the value of $X_{n + 1}$ and applying the Markov property that
\begin{align}
    \mathcal{T}(n; \lambda) &= \sum_{i \in \mathbb{A}^\star} Q_{i} \operatorname{Ex}[\lambda^{\tau_{\mathbb{B}}^{+}(n)} \mid R^+(n + 1),  X_{n} = i] \\
    &= \lambda \sum_{i \in \mathbb{A}^\star}  \sum_{j \in \mathbb{S}} Q_{i} P_{ij}^{+}  \operatorname{Ex}[\lambda^{\tau_{\mathbb{B}}^{+}(n + 1)} \mid R^+(n + 1),  X_{n + 1} = j] \label{eq:Tgf-expected}
\end{align}
We define the following auxiliary generating function that will enable us to obtain an explicit formula for $T(n; \lambda)$: 
\begin{equation}
    V_i(n; \lambda) = \operatorname{Ex}[\lambda^{\tau_{\mathbb{B}}^{+}(n)} \mid R^+(n), X_n = i], \quad i \in \mathbb{S}. 
    \label{eq:Vgen-definition}
\end{equation}
We first compute a system of equations satisfied by $V_i(n; \lambda)$. By definition, we take $V_i(n; \lambda) = 0$ when $i \in \mathbb{A}$ and $V_i(n; \lambda) = 1$ when $i \in \mathbb{B}^\star$. By conditioning on the value of $X_{n + 1}$, we have for $i \notin \mathbb{A}$
\begin{equation} \label{eq:Vgen-expansion}
    V_i(n; \lambda) = \sum_{j \in \mathbb{S}} \operatorname{Pr}(X_{n + 1} = j \mid R^+(n), X_n = i) \operatorname{Ex}[\lambda^{\tau_{\mathbb{B}}^{+}(n)} \mid R^+(n), X_n = i, X_{n + 1} = j].
\end{equation}
When $X_{n + 1} \in \mathbb{B}^\star$, we have $\tau_\mathbb{B}^+(n) = 1$. Otherwise, $\tau_\mathbb{B}^+(n) = 1 + \tau_\mathbb{B}^+(n + 1)$ and the Markov property provide together that
\begin{equation}
    V_i(n; \lambda) = \lambda \left(\sum_{j \in \mathbb{B}^\star}  P_{i, j}^{+} + \sum_{j \in \mathbb{S} \setminus \mathbb{B}^\star}  P_{ij}^{+} \operatorname{Ex}[\lambda^{\tau_{\mathbb{B}}^{+}(n + 1)} \mid R^+(n + 1), X_{n + 1} = j] \right).
\end{equation}
Note that here we have used the fact that $i \notin \mathbb{A}$ to write $\operatorname{Pr}(X_{n + 1} = j \mid R^+(n), X_n = i) = P_{ij}^{+}$. A final application of the Markov property gives $V_i(n; \lambda) = V_i(n + 1;\lambda)$ and hence
\begin{equation} \label{eq:Vgen-recurrence}
    V_i(n; \lambda) = \lambda \left(\sum_{j \in \mathbb{B}^\star}  P_{ij}^{+} +  \sum_{j \in \mathbb{S} \setminus \mathbb{B}^\star} P_{ij}^{+} V_j(n; \lambda) \right).
\end{equation}
Defining the vectors $\mathbf{V}(\lambda) = (V_i(n; \lambda))_{i \in \mathbb{C}}$ and $\mathbf{p} = \smash{\left( \sum_{j \in \mathbb{B}^\star} P_{i, j}^{+} \right)_{i \in \mathbb{C}}}$, as well as the restricted matrix $P^+_{\mathbb{X}\mathbb{Y}} = (P_{ij}^+)_{i \in \mathbb{X}, j \in \mathbb{Y}}$, Eq.~\eqref{eq:Vgen-recurrence} can be written as
\begin{equation} \label{eq:Vgen-system}
    \mathbf{V}(\lambda) = \lambda\left(\mathbf{p} + P_{\mathbb{C}\mathbb{C}}^+\mathbf{V}(\lambda) \right).
\end{equation}
It follows by iteration of Eq.\@~\eqref{eq:Vgen-system} combined with Eq.\@~\eqref{eq:Vgen-definition} that
\begin{equation} \label{eq:Vgen-explicit}
    \mathbf{V}(\lambda) = \sum_{\ell = 1}^{\infty} \lambda^\ell\left(P_{\mathbb{C}\mathbb{C}}^+ \right)^{\ell - 1} \mathbf{p}.
\end{equation}
Returning to Eq.~\eqref{eq:Tgf-expected}, we have
\begin{align} \label{eq:Tgen-in-terms-of-Vgen}
    \mathcal{T}(n; \lambda) &= \lambda \sum_{i \in \mathbb{A}^\star} \sum_{j \in \mathbb{S}} Q_{i} P_{ij}^{+}  V_j(n; \lambda).
\end{align}
Let $\mathbf{Q} = (Q_i)_{i \in \mathbb{A}^\star}$ and $\mathbf{b}^\star = (1)_{i \in \mathbb{B}^\star}$. Collecting powers of $\lambda$ in Eq.\@~\eqref{eq:Tgen-in-terms-of-Vgen}, referring to Eq.\@~\eqref{eq:T-gf-definition}, and applying Eq.~\eqref{eq:Vgen-explicit} gives the desired distribution:
\begin{equation}
    C^{\mathbb{A} \mathbb{B}}(\ell \Delta t) = \begin{cases}
        \mathbf{Q}^\top P^+_{\mathbb{A}^\star\mathbb{B}^\star}\mathbf{b}^\star & \ell = 1 \\ 
        \mathbf{Q}^\top P^+_{\mathbb{A}^\star\mathbb{C}} \left(P^+_{\mathbb{C}\mathbb{C}} \right)^{\ell - 2} P^+_{\mathbb{C}\mathbb{B}^\star} \mathbf{b}^\star & \ell \geq 2,
    \end{cases}
    \label{eq:CAB}
\end{equation}
where we have used the fact that $V_i(n; \lambda) = 1$ when $i \in \mathbb{B}^\star$ to identify the $\ell = 1$ case.

\bibliographystyle{mybst}

\begin{thebibliography}{48}
\expandafter\ifx\csname natexlab\endcsname\relax\def\natexlab#1{#1}\fi

\bibitem[Androulidakis {\rm et~al.}(2016)Androulidakis, Kourafalou, Halliwell, Henaff, Kang, Mehari and Atlas]{Androulidakis-etal-16}
{\rm Androulidakis, Y., Kourafalou, V., Halliwell, G., Henaff, M.~L., Kang, H., Mehari, M. and Atlas, R.} (2016).  {Hurricane interaction with the upper ocean in the Amazon--Orinoco plume region}. {\em Ocean Dynamics\/} 66, 1559--1588.

\bibitem[Balaguru {\rm et~al.}(2012)Balaguru, Chang, Saravanan, Leung, Xu, Li and Hsieh]{Balaguru-etal-12}
{\rm Balaguru, K., Chang, P., Saravanan, R., Leung, L.~R., Xu, Z., Li, M. and Hsieh, J.-S.} (2012).  Ocean barrier layers’ effect on tropical cyclone intensification. {\em Proceedings of the National Academy of Sciences\/} 109~(36), 14343--14347.

\bibitem[Balaguru {\rm et~al.}(2020)Balaguru, Foltz, Leung, Kaplan, Xu, Reul and Chapron]{Balaguru-etal-20}
{\rm Balaguru, K., Foltz, G.~R., Leung, L.~R., Kaplan, J., Xu, W., Reul, N. and Chapron, B.} (2020).  Pronounced impact of salinity on rapidly intensifying tropical cyclones. {\em Bulletin of the American Meteorological Society\/} 101~(9), E1497 -- E1511.

\bibitem[Bonner {\rm et~al.}(2023)Bonner, Beron-Vera and Olascoaga]{Bonner-etal-23}
{\rm Bonner, G., Beron-Vera, F.~J. and Olascoaga, M.~J.} (2023).  Improving the stability of temporal statistics in transition path theory with sparse data. {\em Chaos\/} 33, 063141.

\bibitem[de~Boyer~Montegut {\rm et~al.}(2007)de~Boyer~Montegut, Mignot, Lazar and Cravatte]{deBoyer-etal-07}
{\rm de~Boyer~Montegut, C., Mignot, J., Lazar, A. and Cravatte, S.} (2007).  Control of salinity on the mixed layer depth in the world ocean: 1. {G}eneral description. {\em Journal of Geophysical Research: Oceans\/} 112, C06011.

\bibitem[Br{\'e}maud(1999)]{Bremaud-99}
{\rm Br{\'e}maud, P.} (1999).  {\em Markov chains\/}, vol.~31 of {\em Gibbs Fields Monte Carlo Simulation Queues, Texts in Applied Mathematics\/}. New York: Springer.

\bibitem[Corredor and Morell(2001)]{Corredor-Morell-01}
{\rm Corredor, J.~E. and Morell, J.~M.} (2001).  {Seasonal variation of physical and biogeochemical features in eastern Caribbean Surface Water}. {\em Journal of Geophysical Research\/} 106(C3), 4517--4525.

\bibitem[Domingues {\rm et~al.}(2015)Domingues, Goni, Bringas, Lee, Kim, Halliwell, Dong, Morell and Pomales]{Domingues-etal-15}
{\rm Domingues, R., Goni, G., Bringas, F., Lee, S.-K., Kim, H.-S., Halliwell, G., Dong, J., Morell, J. and Pomales, L.} (2015).  Upper ocean response to hurricane gonzalo (2014): Salinity effects revealed by targeted and sustained underwater glider observations. {\em Geophysical Research Letters\/} 42~(17), 7131--7138.

\bibitem[Domingues {\rm et~al.}(2021)Domingues, Le~Hénaff, Halliwell, Zhang, Bringas, Chardon, Kim, Morell and Goni]{Domingues-etal-21}
{\rm Domingues, R., Le~Hénaff, M., Halliwell, G., Zhang, J.~A., Bringas, F., Chardon, P., Kim, H.-S., Morell, J. and Goni, G.} (2021).  Ocean conditions and the intensification of three major atlantic hurricanes in 2017. {\em Monthly Weather Review\/} 149~(5), 1265 -- 1286.

\bibitem[{E} and {Vanden-Eijnden}(2006)]{E-VandenEijnden-06}
{\rm {E}, W. and {Vanden-Eijnden}, E.} (2006).  Towards a theory of transition paths. {\em J. Stat. Phys.\/} 123, 503--623.

\bibitem[Finkel {\rm et~al.}(2021)Finkel, Webber, Gerber, Abbot and Weare]{Finkel-etal-21}
{\rm Finkel, J., Webber, R.~J., Gerber, E.~P., Abbot, D.~S. and Weare, J.} (2021).  Learning forecasts of rare stratospheric transitions from short simulations. {\em Monthly Weather Review\/} 149, 3647--3669.

\bibitem[Garner(2023)]{Garner-etal-23}
{\rm Garner, A.~J.} (2023).  {Observed increases in North Atlantic tropical cyclone peak intensification rates}. {\em Scientific Reports\/} 13~(1), 16299.

\bibitem[Godfrey and Lindstrom(1989)]{Godfrey-Lindstrom-89}
{\rm Godfrey, J.~S. and Lindstrom, E.~J.} (1989).  {The heat budget of the equatorial western Pacific surface mixed layer}. {\em Journal of Geophysical Research: Oceans\/} 94~(C6), 8007–8017.

\bibitem[Goni and {Tri\~nanes}(2003)]{Goni-Trinanes-03}
{\rm Goni, G.~J. and {Tri\~nanes}, J.~A.} (2003).  {Ocean thermal structure monitoring could aid in the intensity forecast of tropical cyclones}. {\em EOS, Transactions, American Geophysical Union\/} 84, 573--578.

\bibitem[Grodsky {\rm et~al.}(2012)Grodsky, Reul, Lagerloef, Reverdin, Carton, Chapron, Quilfen, Kudryavtsev and Kao]{Grodsky-etal-12}
{\rm Grodsky, S.~A., Reul, N., Lagerloef, G., Reverdin, G., Carton, J.~A., Chapron, B., Quilfen, Y., Kudryavtsev, V.~N. and Kao, H.~Y.} (2012).  {Haline hurricane wake in the Amazon/Orinoco plume:AQUARIUS/SACD and SMOS observations.} {\em Geophysical Research Letters\/} 39.

\bibitem[Halliwell {\rm et~al.}(2015)Halliwell, Gopalakrishnan, Marks and Willey]{Halliwell-etal-15}
{\rm Halliwell, G.~R., Gopalakrishnan, S., Marks, F. and Willey, D.} (2015).  Idealized study of ocean impacts on tropical cyclone intensity forecasts. {\em Monthly Weather Review\/} 143~(4), 1142–1165.

\bibitem[Helfmann(2020)]{Helfmann-20}
{\rm Helfmann, L.} (2020).  Extending transition path theory for time-dependent dynamics. PhD thesis, Free University of Berlin.

\bibitem[Helfmann {\rm et~al.}(2020)Helfmann, Borrell, Sch\"utte and Koltai]{Helfmann-etal-20}
{\rm Helfmann, L., Borrell, E.~R., Sch\"utte, C. and Koltai, P.} (2020).  Extending transition path theory: {P}eriodically driven and finite-time dynamics. {\em J. Nonlinear Sci.\/} 30, 3321--3366.

\bibitem[Hern\'andez {\rm et~al.}(2016)Hern\'andez, Jouanno and Durand]{Hernandez-etal-16}
{\rm Hern\'andez, O., Jouanno, J. and Durand, F.} (2016).  {Do the Amazon and Orinoco freshwater plumes really matter for hurricane‐induced ocean surface cooling?} {\em Journal of Geophysical Research: Oceans\/} 121, 2119--2141.

\bibitem[Hlywiak and Nolan(2019)]{Hlywiak-Nolan-19}
{\rm Hlywiak, J. and Nolan, D.~S.} (2019).  The influence of oceanic barrier layers on tropical cyclone intensity as determined through idealized, coupled numerical simulations. {\em Journal of Physical Oceanography\/} 49~(7), 1723–1745.

\bibitem[Kantha(2006)]{Kantha-06}
{\rm Kantha, L.} (2006).  {Time to replace the Saffir‐-Simpson hurricane scale?} {\em Eos, Transactions American Geophysical Union\/} 87, 3--6.

\bibitem[Landsea and Franklin(2013)]{Landsea-Franklin-13}
{\rm Landsea, W.~C. and Franklin, J.~L.} (2013).  Atlantic hurricane database uncertainty and presentation of a new database format. {\em Mon. Wea. Rev.\/} 141, 3576--3592.

\bibitem[Lasota and Mackey(1994)]{Lasota-Mackey-94}
{\rm Lasota, A. and Mackey, M.~C.} (1994).  {\em Chaos, Fractals and Noise: Stochastic Aspects of Dynamics\/}, 2nd edn., vol.~97 of {\em Applied Mathematical Sciences\/}. New York: Springer.

\bibitem[Lee {\rm et~al.}(2016)Lee, Tippett, Sobel and Camargo]{Lee-etal-16}
{\rm Lee, C.~Y., Tippett, M.~K., Sobel, A.~H. and Camargo, S.} (2016).  Rapid intensification and the bimodal distribution of tropical cyclone intensity. {\em Nat. Commun.\/} 7, 10625.

\bibitem[Lentz(1995)]{Lentz-95}
{\rm Lentz, S.~J.} (1995).  {Seasonal variations in the horizontal structure of the Amazon Plume inferred from historical hydrographic data}. {\em Journal of Geophysical Research\/} 100(C2), 2391–2400.

\bibitem[McDougall and Barker(2011)]{McDougal-Barker-11}
{\rm McDougall, P.~J. and Barker, P.~M.} (2011).  Getting started with teos-10 and the gibbs seawater (gsw) oceanographic toolbox. 28pp., SCOR/IAPSO WG127, ISBN 978-0-646-55621-5.

\bibitem[Metzner {\rm et~al.}(2006)Metzner, {Sch\"utte} and {Vanden-Eijnden}]{Metzner-etal-06}
{\rm Metzner, P., {Sch\"utte}, C. and {Vanden-Eijnden}, E.} (2006).  Illustration of transition path theory on a collection of simple examples. {\em J. Chem. Phys.\/} 125, 084110.

\bibitem[Metzner {\rm et~al.}(2009)Metzner, Sch\"utte and Vanden-Eijnden]{Metzner-etal-09}
{\rm Metzner, P., Sch\"utte, C. and Vanden-Eijnden, E.} (2009).  Transition path theory for {M}arkov jump processes. {\em Multiscale Modeling \& Simulation\/} 7, 1192--1219.

\bibitem[Mignot {\rm et~al.}(2007)Mignot, de~Boyer~Montégut, Lazar and Cravatte]{Mignot-etal-07}
{\rm Mignot, J., de~Boyer~Montégut, C., Lazar, A. and Cravatte, S.} (2007).  {Control of salinity on the mixed layer depth in the world ocean: 2. Tropical areas}. {\em J. Geophys. Res. Oceans\/} 112.

\bibitem[Millero {\rm et~al.}(2008)Millero, Feistel, Wright and McDougall]{Millero-etal-08}
{\rm Millero, F.~J., Feistel, R., Wright, D.~G. and McDougall, T.~J.} (2008).  The composition of standard seawater and the definition of the reference-composition salinity scale. {\em Deep-Sea Res.\/} 55, 50--72.

\bibitem[Miron {\rm et~al.}(2021)Miron, Beron-Vera, Helfmann and Koltai]{Miron-etal-21-Chaos}
{\rm Miron, P., Beron-Vera, F.~J., Helfmann, L. and Koltai, P.} (2021).  Transition paths of marine debris and the stability of the garbage patches. {\em Chaos\/} 31, 033101.

\bibitem[Miron {\rm et~al.}(2019{\natexlab{{\rm a}}})Miron, Beron-Vera, Olascoaga, Froyland, P\'erez-Brunius and Sheinbaum]{Miron-etal-19-JPO}
{\rm Miron, P., Beron-Vera, F.~J., Olascoaga, M.~J., Froyland, G., P\'erez-Brunius, P. and Sheinbaum, J.} (2019{\natexlab{{\rm a}}}).  {Lagrangian geography of the deep Gulf of Mexico}. {\em J. Phys. Oceanogr.\/} 49, 269--290.

\bibitem[Miron {\rm et~al.}(2019{\natexlab{{\rm b}}})Miron, Beron-Vera, Olascoaga and Koltai]{Miron-etal-19-Chaos}
{\rm Miron, P., Beron-Vera, F.~J., Olascoaga, M.~J. and Koltai, P.} (2019{\natexlab{{\rm b}}}).  {Markov-chain-inspired search for MH370}. {\em Chaos: An Interdisciplinary Journal of Nonlinear Science\/} 29, 041105.

\bibitem[Muller-Karger {\rm et~al.}(1988)Muller-Karger, McClain and Richardson]{Muller-etal-88}
{\rm Muller-Karger, F.~E., McClain, C.~R. and Richardson, P.} (1988).  {The dispersal of Amazon's water}. {\em Nature\/} 333, 56--59.

\bibitem[Neetu {\rm et~al.}(2012)Neetu, Lengaigne, Vincent, Vialard, Madec, Samson, Kumar and Durand]{Neetu-etal-12}
{\rm Neetu, S., Lengaigne, M., Vincent, E., Vialard, J., Madec, G., Samson, G., Kumar, M. R.~R. and Durand, F.} (2012).  Influence of upper-ocean stratification on tropical cyclone-induced surface cooling in the bay of bengal. {\em Journal of Geophysical Research: Oceans\/} 117~(C12).

\bibitem[Norris(1998)]{Norris-98}
{\rm Norris, J.} (1998).  {\em Markov Chains\/}. Cambridge University Press.

\bibitem[Rudzin {\rm et~al.}(2019)Rudzin, Shay and Jaimes]{Rudzin-etal-19}
{\rm Rudzin, J.~E., Shay, L.~K. and Jaimes, B.} (2019).  {UThe Impact Of The Amazon–Orinoco River Plume On Enthalpy Flux And Air–Sea Interaction Within Caribbean Sea Tropical Cyclones}. {\em Monthly Weather Review\/} 147.

\bibitem[Sch\"utte {\rm et~al.}(2016)Sch\"utte, Koltai and Klus]{Schutte-etal-16}
{\rm Sch\"utte, C., Koltai, P. and Klus, S.} (2016).  On the numerical approximation of the {Perron--Frobenius} and {Koopman} operator. {\em Journal of Computational Dynamics\/} 3~(1), 1–12.

\bibitem[Seroka {\rm et~al.}(2017)Seroka, Miles, Xu, Kohut, Schofield and Glenn]{Seroka-etal-17}
{\rm Seroka, G., Miles, T., Xu, Y., Kohut, J., Schofield, O. and Glenn, S.} (2017).  Rapid shelf-wide cooling response of a stratified coastal ocean to hurricanes. {\em Journal of Geophysical Research: Oceans\/} 122~(6), 4845--4867.

\bibitem[Shay {\rm et~al.}(2000)Shay, Goni and Black]{Shay-etal-00}
{\rm Shay, L.~K., Goni, G.~J. and Black, P.~G.} (2000).  Effects of a warm oceanic feature on hurricane opal. {\em Mon. Weather Rev.\/} 128, 1366--1383.

\bibitem[Sloyan {\rm et~al.}(2019)Sloyan, Wanninkhof, Kramp, Johnson, Talley, Tanhua, McDonagh, Cusack, O’Rourke, McGovern, Katsumata, Diggs, Hummon, Ishii, Azetsu-Scott, Boss, Ansorge, Perez, Mercier, Williams, Anderson, Lee, Murata, Kouketsu, Jeansson, Hoppema and Campos]{Sloyan-etal-19}
{\rm Sloyan, B.~M., Wanninkhof, R., Kramp, M., Johnson, G.~C., Talley, L.~D., Tanhua, T., McDonagh, E., Cusack, C., O’Rourke, E., McGovern, E., Katsumata, K., Diggs, S., Hummon, J., Ishii, M., Azetsu-Scott, K., Boss, E., Ansorge, I., Perez, F.~F., Mercier, H., Williams, M. J.~M., Anderson, L., Lee, J.~H., Murata, A., Kouketsu, S., Jeansson, E., Hoppema, M. and Campos, E.} (2019).  The global ocean ship-based hydrographic investigations program (go-ship): A platform for integrated multidisciplinary ocean science. {\em Frontiers in Marine Science\/} 6.

\bibitem[Tarjan(1972)]{Tarjan-72}
{\rm Tarjan, R.} (1972).  Depth-first search and linear graph algorithms. {\em SIAM J. Comput.\/} 1, 146--160.

\bibitem[Vanden-Eijnden(2006)]{VandenEijnden-06}
{\rm Vanden-Eijnden, E.} (2006).  {\em Transition Path Theory\/}, p. 453–493. Springer Berlin Heidelberg.

\bibitem[Wang {\rm et~al.}(2011)Wang, Han, Qi and Li]{Wang-etal-11}
{\rm Wang, X.~D., Han, G.~J., Qi, Y.~Q. and Li, W.} (2011).  Impact of barrier layer on typhoon-induced sea surface cooling. {\em Dynamics of Atmospheres and Oceans\/} 52~(3), 367--385.

\bibitem[Yan {\rm et~al.}(2017)Yan, Li and Wang]{Yan-etal-17}
{\rm Yan, Y., Li, L. and Wang, C.} (2017).  The effects of oceanic barrier layer on the upper ocean response to tropical cyclones. {\em Journal of Geophysical Research: Oceans\/} 122, 4829--4844.

\bibitem[Zhang {\rm et~al.}(2022)Zhang, Ma, Fei, Zheng and Huang]{Zhang-etal-22}
{\rm Zhang, Z., Ma, Z., Fei, J., Zheng, Y. and Huang, J.} (2022).  The effects of tropical cyclones on characteristics of barrier layer thickness. {\em Frontiers in Earth Science\/} 10, 962232.

\bibitem[Zuo {\rm et~al.}(2015)Zuo, Balmaseda and Mogensen]{Zuo-etal-15}
{\rm Zuo, H., Balmaseda, M.~A. and Mogensen, K.} (2015).  {The new eddy-permitting ORAP5 ocean reanalysis: description, evaluation and uncertainties in climate signals}. {\em Climate Dynamics\/} 49, 791–811.

\bibitem[Zuo {\rm et~al.}(2019)Zuo, Balmaseda, Tietsche, Mogensen and Mayer]{Zuo-etal-19}
{\rm Zuo, H., Balmaseda, M.~A., Tietsche, S., Mogensen, K. and Mayer, M.} (2019).  {The ECMWF operational ensemble reanalysis–analysis system for ocean and sea ice: a description of the system and assessment}. {\em Ocean Science\/} 15, 779–808.

\end{thebibliography}

\end{document}